\documentclass{article}

\usepackage{arxiv}

\usepackage{times}
\usepackage{graphicx}
\usepackage{amssymb}

\usepackage{authblk}

\usepackage{xurl} 
\usepackage[caption=false,font={scriptsize,sf}]{subfig}
\usepackage{hyperref}
\usepackage{multirow}
\usepackage{xfrac}
\usepackage[shortcuts]{extdash} % then write containers\-/as\-/a\-/service if needed
\usepackage{booktabs}
\usepackage{threeparttable}
\usepackage{tabularx}
\usepackage{tabto}
\usepackage[svgnames]{xcolor}
\usepackage{txfonts}
\newcommand*{\graybullet}{\textcolor{gray!40}{\medbullet}}
\usepackage{relsize}

\hypersetup{
    colorlinks=true,
    urlcolor=blue,
    citecolor=blue,
    linkcolor=blue,
}

\begin{document}

\renewcommand{\UrlFont}{\rmfamily\smaller}

\title{Multitenant Containers as a Service \\
(CaaS) for Clouds and Edge Clouds}

%\date{September 9, 1985}	% Here you can change the date presented in the paper title
%\date{} 					% Or removing it

\author[1,2]{Berat~Can~{\c{S}}enel}
\author[1,2]{Maxime~Mouchet}
\author[3]{Justin~Cappos}
\author[1]{Olivier~Fourmaux}
\author[1,2]{Timur~Friedman}
\author[4]{Rick~McGeer}
\affil[1]{\textsc{Lip}6-CNRS lab, Sorbonne Université}
\affil[2]{\textsc{Lincs} lab}
\affil[3]{NYU Tandon School of Engineering}
\affil[4]{engageLively}

% Uncomment to remove the date
%\date{}

% Uncomment to override  the `A preprint' in the header
\renewcommand{\headeright}{Preprint}
\renewcommand{\undertitle}{Preprint \\ This work has been submitted to the IEEE for possible publication. Copyright may be transferred without notice, after which this version may no longer be accessible.}
\renewcommand{\shorttitle}{Multitenant CaaS for Clouds and Edge Clouds}

%%% Add PDF metadata to help others organize their library
%%% Once the PDF is generated, you can check the metadata with
%%% $ pdfinfo template.pdf
\hypersetup{
pdftitle={Multitenant CaaS for Clouds and Edge Clouds},
pdfsubject={cs.DC, cs.OH},
pdfauthor={Berat~Can~{\c{S}}enel, Maxime~Mouchet, Justin~Cappos, Olivier~Fourmaux, Timur~Friedman, Rick~McGeer},
pdfkeywords={Edge computing, Cloud computing, Containers as a Service, Multitenancy, Federation, Kubernetes},
}

\maketitle
\thispagestyle{empty}

\begin{abstract}

Cloud computing, offering on-demand access to computing resources through the Internet and the pay-as-you-go model, has marked the last decade with its three main service models; Infrastructure as a Service (IaaS), Platform as a Service (PaaS), and Software as a Service (SaaS).
The lightweight nature of containers compared to virtual machines has led to the rapid uptake of another in recent years, called Containers as a Service (CaaS), which falls between IaaS and PaaS regarding control abstraction.
However, when CaaS is offered to multiple independent users, or tenants, a multi-instance approach is used, in which each tenant receives its own separate cluster, which reimposes significant overhead due to employing virtual machines for isolation.
If CaaS is to be offered not just at the cloud, but also at the edge cloud, where resources are limited, another solution is required.  
We introduce a native CaaS multitenancy framework, meaning that tenants share a cluster, which is more efficient than the one tenant per cluster model.
Whenever there are shared resources, isolation of multitenant workloads is an issue.
Such workloads can be isolated by Kata Containers today.
Besides, our framework esteems the application requirements that compel complete isolation and a fully customized environment.
Node-level slicing empowers tenants to programmatically reserve isolated subclusters where they can choose the container runtime that suits application needs.
The framework is publicly available as liberally-licensed, free, open-source software that extends Kubernetes, the de facto standard container orchestration system.
It is in production use within the EdgeNet testbed for researchers.

\end{abstract}

% keywords can be removed

\section{Introduction}
\label{sec-introduction}
Multitenancy is what makes cloud computing economical.
From a single bare metal machine, a cloud provider can offer resources to multiple tenants, where each tenant is a customer that contracts for cloud services on behalf of one or more users.
These resources are, for example, virtual machines in the Infrastructure as a Service (IaaS) service model, or tools for application development and deployment in the Platform as a Service (PaaS) model.
Tenants that are prepared to accept less than perfect isolation from other tenants benefit from the lower prices that providers can offer thanks to more efficient use of the providers' hardware.

But, despite the greater efficiency of containers as compared to virtual machines, and despite recent improvements in ensuring isolation between containers, the cloud industry does not yet propose a multitenant Containers as a Service (CaaS) offering that takes advantage of these advances.  
What passes for CaaS today is in fact multiple side-by-side instances of single-tenant clusters of compute nodes, each cluster having its own container orchestration control plane and its own data plane, and isolated from other clusters through the use of virtual machines.
For example, automated services such as AWS Fargate\footnote{Amazon Web Services' Fargate \url{https://aws.amazon.com/fargate}} and Google Autopilot\footnote{Google Cloud's Autopilot \url{https://cloud.google.com/kubernetes-engine/docs/concepts/autopilot-overview}} that manage cluster capacity on behalf of a user who is deploying containers to the cloud do not do away with virtual machine overhead and do not improve control plane efficiency.\footnote{Google Cloud documentation: Cluster Architecture \url{https://cloud.google.com/kubernetes-engine/docs/concepts/cluster-architecture\#nodes}}
In brief, although CaaS ought to offer greater efficiency than IaaS,\footnote{We make the assumption that IaaS is offered through virtual machines, which is commonly the case~\cite{felter2015containervsvm}.} it does not yet do so.

With the emergence of the edge cloud, such efficiency will take on greater importance because resources will typically be more constrained than in the cloud.
As part of the vision for 5G, it is projected that mobile network operators will become edge cloud providers, offering up compute resources from servers that are colocated with their wireless base stations~\cite{ding2018apps5g, liyanage2021MECIoT}, at what is being termed the `service provider edge'~\cite{linux2020whitepaper, linux2021report, linux2022report}.
These operators are also expected to offer resources from their peering sites, or the `regional edge'~\cite{linux2022whitepaper}.
Such edge cloud instances will be data centers that are geographically dispersed to be closer to the users of cloud services or to edge devices than are the centralized data centers that dominate the present-day cloud.\footnote{To be clear, we do not include low-powered IoT devices that are unable to run cloud-like workloads (the `constrained device edge')~\cite{linux2020whitepaper} in our conception of the edge cloud that we anticipate for CaaS.}
With fewer resources, an edge cloud will not scale as elastically as a cloud, yet it must be prepared to receive a large number of workloads that have been deployed to serve local users and devices. 

The problem that we aim to resolve is how to move CaaS multitenancy away from a high-overhead multi-instance model to a more efficient one that will be suitable for the resource-constrained edge cloud. 
In the solution that we propose, multiple tenants share a single instance of the control plane, which is used to deploy containers that coexist within a single instance of a shared cluster, while still allowing tenants to enjoy isolation from each other as well as the opportunity to customize their resources.

Our multitenancy solution has the particularity that it is designed to work in a federated environment.
Today, a cloud customer typically deploys their workloads to a single cloud provider, but if they want to extend those workloads to be close to users and edge devices, a customer will also need to obtain resources from multiple edge cloud providers~\cite{chiang2016fog}.\footnote{In addition, a customer might bring resources to bear from its own `user edge'.}
Doing so will be easiest for a customer if those providers are federated, meaning that the customer will be able to contract with just one cloud or edge cloud provider and the customer will be able to deploy its workloads through a single interface offered by that provider~\cite{yi2015fogsurvey, javed2020iotef}, and the provider will manage the propagation of the workloads to the other providers.
Accordingly, our multitenancy solution ensures that each cloud provider can accept tenant workloads that originate from other providers.

As we use the term, a \textit{multitenancy framework} consists of a set of rules that govern how a cloud provider offers resources to its tenants such that each tenant can use their portion of the resources and configure those resources to meet their needs without regard for the presence of the other tenants.
The rules address the creation of isolated environments, resource sharing, and user permission management.
They determine which rights over resources are given to which tenants, under which conditions, and how those rights affect the relationships of other tenants with the same resources.
The term equally well refers to the set of entities that are coded to enforce these rules.

In this paper, we describe our framework, argue for it, and show how we have implemented it in EdgeNet, a production edge cloud.\footnote{The EdgeNet testbed \url{https://edge-net.org/}} 
What we henceforth refer to as the \textit{EdgeNet multitenancy framework} is part of the larger EdgeNet code base,\footnote{The EdgeNet software \url{https://github.com/EdgeNet-project/edgenet}} which is free, liberally-licensed, and open source software that enables CaaS deployments to the edge cloud.
It is desiged as a set of extensions to the Kubernetes container orchestration system,\footnote{Kubernetes \url{https://github.com/kubernetes/kubernetes}} which is itself free, liberally-licensed, and open source.
Our reasoning in building upon Kubernetes is that cloud customers will want to continue using this familiar system, which is today's de facto industry standard container orchestration tool.

As Kubernetes does not natively support multitenancy, others have identified the need for such an extension and have developed their own Kubernetes multitenancy frameworks. 
(See Table~\ref{tab:related-work-comparison} for details.)
We will show that the existing frameworks, while no doubt fine for the cloud, will not be suitable for CaaS in the edge cloud.
There are a few prior studies concerning these frameworks~\cite{zheng2021multi, goethals2022virtualkubelets, dubey2022usermanagement}, but this is the first paper to situate them, and EdgeNet, within the existing scientific literature on cloud multitenancy.

Our contributions, and the sections of the paper that address them, are as follows:
\begin{itemize}
    \item \noindent We look at Kubernetes multitenancy frameworks through the lens of the scientific literature on cloud multitenancy and, in Sec.~\ref{sec-multitenancy-approach}, we provide a novel classification of these frameworks into three main approaches: multi-instance through multiple clusters, multi-instance through multiple control planes, and single-instance native.
    \item \noindent Based upon our analysis of the literature, we distill out four features that we believe will promote a future in which CaaS can thrive, in particular at the network edge, and we describe how we have incorporated these features into the EdgeNet multitenancy framework: consumer and vendor tenancy in Sec.~\ref{sec-consumer-and-vendor-modes}, tenant resource quota for hierarchical namespaces in Sec.~\ref{sec-tenant-resource-quota}, variable slice granularity in Sec.~\ref{sec-variable-slice-granularity}, and federation support in Sec.~\ref{sec-federation-support}.
    \item \noindent We have implemented the EdgeNet multitenancy framework as a free and open-source extension to Kubernetes, and have put it into production as the EdgeNet testbed, as described in Sec.~\ref{sec-architecture}.
    \item \noindent Our EdgeNet multitenancy framework constitutes a prototype for the federation of clouds and edge clouds, and we provide a vision in Sec.~\ref{sec-architecture-subnamespace-federation} for the future development of a full federation framework.
    \item \noindent We benchmark the three multitenancy framework approaches using a representative implementation for each approach, and we reveal their pros and cons from a tenancy-centered edge computing perspective in Sec.~\ref{sec-benchmarking}.
\end{itemize}

The paper is structured as follows. 
Sec.~\ref{sec-rationale} provides background on cloud multitenancy, the challenges that it presents, and the ways in which those challenges have been addressed for edge computing.
Sec.~\ref{sec-related-work} describes related work in the specific area of Kubernetes multitenancy frameworks.
Sec.~\ref{sec-design} discusses  design principles for a CaaS multitenancy framework, and Sec.~\ref{sec-architecture} presents the architecture of the EdgeNet multitenancy framework that we have developed.
In Sec.~\ref{sec-benchmarking}, we benchmark our framework against representative frameworks for two alternate approaches, and we point to our future work in Sec.~\ref{sec-future-work}.

\section{Rationale}
\label{sec-rationale}
We envisage a future in which tenants deploy services on a continuum of computing resources from cloud to edge cloud, about which we make the following assumptions:
\begin{itemize}
    \item Edge clouds are ubiquitous, scattered across the world~\cite{ding2020cloudedge}.
    \item Compute and storage resources are constrained in the edge cloud, making it harder to scale tenant workloads there than in the cloud.
    \item Tenants value the ability to easily move their workloads from one edge cloud cluster to another and between the edge cloud and the cloud.
    \item Each tenant's user database is maintained by that tenant. User management is not a functionality provided by the compute clusters. 
    \item Tenants and their users are unreliable.
    They may purposely or accidentally harm each other, or the compute cluster, or themselves.
\end{itemize}

We conceive of our proposed architecture based on these assumptions, for which we provide rationale in the following subsections: the necessity of a novel Kubernetes CaaS multitenancy framework (Sec.~\ref{sec-rationale-multi-tenancy}) that takes container-specific security and performance considerations into account (Sec.~\ref{sec-rationale-security-and-performance}), and that enables federation across edge clouds and control over slice granularity at the edge (Sec.~\ref{sec-rationale-edge-computing-federation-and-slicing}).

\subsection{Multitenancy}
\label{sec-rationale-multi-tenancy}

It is an often-repeated commonplace that cloud computing is not just ``using someone else's computer'', as the cloud goes beyond this to promise more flexible, convenient, and cost-effective access to computing resources.
Multitenancy is required to realize this promise.
The \textit{NIST Definition of Cloud Computing}~\cite{mell2011nist} mentions resource pooling as one of the ``five essential characteristics'' of cloud computing, saying that:
\begin{quote}
    \textit{The provider’s computing resources are pooled to serve multiple consumers using a multi-tenant model, with different physical and virtual resources dynamically assigned and reassigned according to consumer demand.}
\end{quote}
\noindent And in their \textit{Defining Multi-Tenancy} paper from 2014~\cite{KABBEDIJK2015139}, Kabbedijk et al.\@ state:
\begin{quote}
    \textit{Multi-tenancy is a property of a system where multiple customers, so-called tenants, transparently share the system’s resources, such as services, applications, databases, or hardware, with the aim of lowering costs, while still being able to exclusively configure the system to the needs of the tenant.}
\end{quote}

Multitenancy is a standard feature of the three established cloud service models, Software as a Service (SaaS), Platform as a Service (PaaS), and Infrastructure as a Service (IaaS)~\cite{rimal2011architectural, bien2014hierarchical}.
If CaaS is to provide the promised benefits of the cloud and the edge cloud at scale, then it requires an efficient multitenancy model as well.
We further discuss why such efficiency is required for CaaS to run for clouds and edge clouds in Sec.~\ref{sec-multitenancy-approach}, and the results of our experiments in Sec.~\ref{sec-benchmarking} support our contention.

Multitenancy has a broad meaning and can be enabled at different cloud abstraction layers using different techniques to share resources among multiple customers.
This paper discusses multitenancy in the context of CaaS and methods for accomplishing it.
CaaS offerings are mostly based upon Kubernetes~\cite{zheng2021multi}, so we focus on the ways in which it can serve multiple customers using multitenancy.
To be clear with respect to the discussion of multitenancy in the Kubernetes documentation,\footnote{Kubernetes documentation: \textit{Multi-tenancy} \url{https://kubernetes.io/docs/concepts/security/multi-tenancy/}} which describes how a tenant can deploy an application in a Kubernetes cluster to serve its multiple customers using a multi-tenant model: that is also multitenancy, but at the application layer, and more precisely at the SaaS layer, however it is not multi-tenant CaaS, which is what this paper considers.

\subsection{Security and Performance}
\label{sec-rationale-security-and-performance}

While multitenancy is an essential cloud feature, it raises security issues that researchers have been considering for over a decade~\cite{buyya2011platforms}, notably with respect to the IaaS service model~\cite{dawoud2010infrastructure}.
For example, potential users are concerned about the security of their data when multiple tenants share the same infrastructure~\cite{benlian2011opportunities}, and the resulting lack of trust can hamper cloud adoption~\cite{rimal2011architectural}.

Virtualization is used to isolate tenants from one another, but containers tend to offer weaker isolation~\cite{bernstein2014containers}, which introduces new concerns for multitenant container platforms~\cite{RODEROMERINO201296}, such as information leakage between colocated containers~\cite{gao2017containerleaks}. In general, Sultan et al.~\cite{sultan2019container} have identified four categories of threat in containerized environments: malicious applications within containers, one container harming another, a container harming its host, and a container within an untrustworthy host.

In Kubernetes, container security must be considered in the context of the \textit{pod}, which is that system's smallest deployable unit, consisting of a set of one or more containers.
The Kubernetes pod security standards define three profiles, Privileged, Baseline, and Restricted.\footnote{Kubernetes documentation: \textit{Cloud Security Standards} \url{https://kubernetes.io/docs/concepts/security/pod-security-standards/}}
However, these standards address a single-tenant environment, and so overlook some of the multitenant security issues mentioned above.

We therefore see the need for a solution that diminishes the security risks of running colocated containerized workloads.
In order to be of interest for CaaS, such a solution needs to maintain the performance advantage of containers over virtual machines.

\subsection{Edge Computing, Federation, and Slicing}
\label{sec-rationale-edge-computing-federation-and-slicing}

As described in the Linux Foundation's 2021 \textit{State of the Edge} report~\cite{linux2021report}, cloud-like infrastructure is being developed at the network edge in order to serve edge devices that produce bandwidth-intensive and/or latency-sensitive workloads.
\textsc{ETSI}'s multi-access edge computing (\textsc{MEC}) architecture~\cite{etsi2020mec} provides a standard structure for making servers at cellular operators' radio access networks available for the deployment of such workloads by third parties.
That is, the emerging edge cloud will be a multitenant cloud~\cite{ananthanarayanan2017edge}.

Since the \textsc{MEC} architecture anticipates that workloads may be containerized, we argue that there is a need for a multitenant CaaS framework that meets the specific requirements of the network edge. 
The prime edge requirements that we identify are federation and variable slice granularity.

\textsc{MEC} facilities will be provided by multiple operators.
Just as a mobile phone user is able to roam from one regional operator to another today, a mobile edge device will need to be able to connect to different operators and find its containerized edge services spun up near each base station to which it connects.
And \textsc{ETSI} describes a requirement for edge devices to be able to engage in low-latency interactions with each other when they are near each other, even if they are connected to different operators' base stations.
\textsc{ETSI} uses the term \textit{federation} to describe such interoperability scenarios.

To enable federation, we argue, a CaaS framework must support the deployment of third parties' containers across multiple operators' edge clouds.
That is, the framework will not just be multitenant, it will also be multi-provider, with providers furnishing geographically dispersed heterogeneous resources.
Those who deploy CaaS services to a multi-provider environment will be in need of a unified interface that simplifies the task of moving workloads between remote clusters that are owned by different providers~\cite{yi2015fogsurvey}.

In addition, as anticipated by the Next Generation Mobile Networks Alliance in 2016~\cite{ngmn2016slicing}, operators will have to support third party services that put a much more heterogeneous set of requirements on their networks than is currently the case.
Extreme requirements are incompatible with a one-size-fits-all approach.
The way that \textsc{MEC} handles this is through \textit{slicing}~\cite{liyanage2021MECIoT,etsi2019slicing,zanzi2018m}, which allows network and compute resources to be allocated and custom-configured to meet the specific needs of individual services.
In the CaaS context, we argue that no single slice granularity will meet the full range of needs.
The standard CaaS sub-node-level slicing, in which containers are provided from a shared resource pool on individual node, while no doubt appropriate for many services, will not be appropriate for those that are the most sensitive to performance variation.
For those services, node-level slice granularity will be needed.

\section{Related Work}
\label{sec-related-work}
Someone who wishes to deploy containerized services to the cloud has a choice of open source container orchestration systems with which to do so, four of the most prominent being~\cite{aljawarneh2019container}: Apache Mesos,\footnote{Apache Mesos \url{https://github.com/apache/mesos}} Docker Swarm,\footnote{Docker Swarm \url{https://github.com/docker/swarmkit}} Kubernetes,\footnote{Kubernetes \url{https://kubernetes.io/}} and Rancher's Cattle.\footnote{Rancher's Cattle \url{https://github.com/rancher/cattle}}
We focus on Kubernetes, as it has in recent years become the de facto industry standard.
All of the major cloud providers offer Kubernetes-based CaaS to their customers (see Table~\ref{tab:commercial-caas}).
And Datadog, a company that provides cloud monitoring and security services, reports~\cite{datadog2022insights} that nearly 50\% of their customers that deploy containers use Kubernetes to do so, this having increased about 10 percentage points over the past three years.

\begin{table*}[t]
\sffamily % a sans serif font is recommended for tables
\scriptsize
\renewcommand{\arraystretch}{1.2} % let the table breathe a bit more
\caption{Major cloud providers' Kubernetes-based containers-as-a-service (CaaS) offerings}
\label{tab:commercial-caas}
\centering
\renewcommand{\UrlFont}{\rmfamily\tiny}
    \begin{tabular}{lll}
        \toprule
        \textbf{Cloud provider} & \textbf{Kubernetes-based CaaS offering} & \textbf{URL} \\
        \midrule
        Amazon Web Services & Elastic Kubernetes Service &  \url{https://aws.amazon.com/eks/} \\
        Microsoft Azure & Azure Kubernetes Service & \url{https://azure.microsoft.com/en-us/products/kubernetes-service} \\
        Google Cloud Platform & Google Kubernetes Engine & \url{https://cloud.google.com/kubernetes-engine} \\
        Alibaba Cloud & Alibaba Cloud Container Service for Kubernetes & \url{https://www.alibabacloud.com/product/kubernetes} \\
        Oracle Cloud & Oracle Container Engine for Kubernetes & \url{https://www.oracle.com/cloud/cloud-native/container-engine-kubernetes/} \\
        IBM Cloud & IBM Cloud Kubernetes Service & \url{https://www.ibm.com/cloud/kubernetes-service} \\
        Tencent Cloud & Tencent Kubernetes Engine & \url{https://www.tencentcloud.com/products/tke} \\
        OVHcloud & Free Managed Kubernetes & \url{https://us.ovhcloud.com/public-cloud/kubernetes/} \\
        DigitalOcean & DigitalOcean Kubernetes & \url{https://try.digitalocean.com/kubernetes-in-minutes/} \\
        Linode & Linode Kubernetes Engine & \url{https://www.linode.com/lp/kubernetes/} \\
        \bottomrule
    \end{tabular}
\end{table*}
\renewcommand{\UrlFont}{\rmfamily\smaller}
In the commercial cloud offerings, each customer gets their own Kubernetes cluster, which is a straightforward form of multitenancy.
Some providers add on more advanced features.
For example, an Amazon EKS customer can use a service called Fargate\footnote{Amazon Web Services (AWS) Elastic Kubernetes Cloud (EKS) documentation: \textit{Fargate} \url{https://docs.aws.amazon.com/eks/latest/userguide/fargate.html}} to manage the capacity of their Kubernetes cluster, adding and removing nodes as they need to.
Similarly, a Google Cloud customer can hand over control of their cluster capacity management to a service called Autopilot,\footnote{Google Cloud documentation: \textit{Create an Autopilot cluster} \url{https://cloud.google.com/kubernetes-engine/docs/how-to/creating-an-autopilot-cluster}} to do the same thing for them automatically.
% Microsoft Azure's AKS.\footnote{Microsoft's Azure Kubernetes Service (AKS) documentation \url{https://learn.microsoft.com/en-us/azure/aks/}}
 
While Kubernetes multitenancy in this form might be fine for large centralized data center clouds, there are drawbacks when looking to an edge cloud future. 
Setting up a separate cluster for each tenant is far from the most efficient approach, as we will show in Sec.~\ref{sec-benchmarking}.
Resources are liable to be underused, which will be of particular concern in the smaller data centers that we can anticipate at the edge.
And when tenants need to be repeatedly instantiated as their workloads migrate, for instance at one roadside cabinet after another to serve vehicles that are moving along a highway, spinning up an entire cluster for each arrival of a tenant risks taking too much time.
We anticipate that lighter forms of multitenancy will be needed: ones that allow more efficient resource sharing, even at some cost in workload isolation, and that allow more rapid creation and deletion of tenants.
Furthermore, proprietary systems for enabling multitenancy risk being a hindrance in a federated environment, in which a single customer might deploy their workloads to many edge clouds, each owned by a different operator.
If all of the operators use a common open-source multitenancy framework, it will promote interoperability.

Starting in 2019, as Table~\ref{tab:related-work-comparison} shows, a fair number of open-source Kubernetes multitenancy frameworks have been developed.
Some, such as Virtual Kubelet~\cite{virtualkubelet2020virtualkubelet} and frameworks that are derived from that code, take the same starting point as the commercial services, which is each tenant having its own cluster.
But others offer worker nodes to tenants out of a shared cluster, which is more resource efficient.
And some of these serve multiple tenants out of a shared control plane, which is yet more efficient.

The Kubernetes community has recognized the importance of developing such frameworks, as evidenced by the fact that one of the Kubernetes working groups, of which there are just five,\footnote{Kubernetes working groups \url{https://github.com/kubernetes/community/blob/master/sig-list.md}} is devoted to multitenancy.\footnote{Kubernetes Multi-tenancy Working Group \url{https://github.com/kubernetes-sigs/multi-tenancy}}
Both of the frameworks that this working group supports take the shared cluster approach.
VirtualCluster (VC)~\cite{virtualcluster2021virtualcluster} offers a separate control plane to each tenant while the control plane is shared among tenants by the Hierarchical Namespace Controller (HNC)~\cite{hnc2019hnc}.
These two frameworks, along with the others shown in Table~\ref{tab:related-work-comparison}, comprise the essential related work for our own EdgeNet framework.

We look at six aspects of Kubernetes multitenancy frameworks when comparing EdgeNet to the related work: the multitenancy approach (Sec.~\ref{sec-multitenancy-approach}), the customization approach (Sec.~\ref{sec-tenant-environment-customization}), support for consumer and vendor modes (Sec.~\ref{sec-consumer-and-vendor-modes}), management of tenant resource quotas (Sec.~\ref{sec-tenant-resource-quota}), support for variable slice granularities (Sec.~\ref{sec-variable-slice-granularity}), and support for federation (Sec.~\ref{sec-federation-support}). 

\newcolumntype{Y}{>{\centering\arraybackslash}X}
\begin{table*}[t]
\sffamily % a sans serif font is recommended for tables
\scriptsize
\centering
\renewcommand{\arraystretch}{1.2} % let the table breathe a bit more
\caption{Comparison table of related work (open-source Kubernetes multitenancy frameworks)}
\label{tab:related-work-comparison}
\begin{threeparttable}
\begin{tabularx}{\linewidth}{l@{\extracolsep{\fill}}YYYYYYYYYY}
    \toprule
    & \textbf{EdgeNet} & \textbf{HNC} & \textbf{Capsule} & \textbf{kiosk} & \textbf{Arktos} & \textbf{VC} & \textbf{k3v} & \textbf{vcluster} & \textbf{Kamaji} & \textbf{VK+} \\
    & v1.0.0-alpha.5 & v1.0.0 & v0.3.1 & v0.2.11 & v1.0 & v0.1.0 & v0.0.1 & v0.15.0-alpha.1 & v0.2.1 & VK v1.8.0 \\    
    & 2023-04 & 2022-04 & 2023-03 & 2021-11 & 2022-03 & 2021-06 & 2019-07 & 2023-03 & 2023-02 & 2023-03 \\
    \midrule
    \multicolumn{11}{l}{\textbf{Multitenancy Approach}} \\
    \multicolumn{11}{l}{\hspace{5pt}Multi-instance} \\ 
    \hspace{10pt}- Through Multiple Clusters & & & & & & & & & & $\graybullet$ \\
    \hspace{10pt}- Through Multiple Control Planes & & & & & & $\graybullet$ & $\graybullet$ & $\graybullet$ & $\graybullet$ & \\ 
    \multicolumn{11}{l}{\hspace{5pt}Single-instance} \\ 
    \hspace{10pt}- Single-instance Native & $\graybullet$ & $\graybullet$ & $\graybullet$ & $\graybullet$ & $\graybullet$ & & & & & \\     
    \midrule
    \multicolumn{11}{l}{\textbf{Customization Approach}} \\ 
    \multicolumn{11}{l}{\hspace{5pt}Control Plane} \\ 
    \hspace{10pt}- Full Control Plane View & & & & & & $\graybullet$ & $\graybullet$ & $\graybullet$ & $\graybullet$ & $\graybullet$ \\ 
    \hspace{10pt}- Tenant-wise Abstraction & & & & & $\graybullet$ & & & & & \\ 
    \hspace{10pt}- Flat Namespaces & & & $\graybullet$ & $\graybullet$ & & & & & & \\ 
    \hspace{10pt}- Hierarchical Namespaces & $\graybullet$ & $\graybullet$ & & & & & & & & \\ 
    \multicolumn{11}{l}{\hspace{5pt}Data Plane} \\ 
    \hspace{10pt}- SSH Access to Worker Nodes & & & & & & & & & $\graybullet$ & Partial \\ 
    \midrule
    \multicolumn{11}{l}{\textbf{Consumer \& Vendor Modes}} \\ 
    \hspace{5pt}- Consumer Mode & $\graybullet$ & $\graybullet$ & $\graybullet$ & $\graybullet$ & $\graybullet$ & $\graybullet$ & $\graybullet$ & $\graybullet$ & $\graybullet$ & $\graybullet$ \\
    \hspace{5pt}- Vendor Mode & $\graybullet$ & & & & & & & & & $\graybullet$ \\
    \midrule
    \textbf{Tenant Resource Quota} & $\graybullet$ & Incomplete & $\graybullet$ & $\graybullet$ & $\graybullet$ & $\graybullet$ & & $\graybullet$ & $\graybullet$ & $\graybullet$ \\
    \midrule
    \multicolumn{11}{l}{\textbf{Variable Slice Granularity}} \\
    \hspace{5pt}- Node-level Slicing & $\graybullet$ & $\graybullet$ & $\graybullet$ & $\graybullet$ & $\graybullet$ & $\graybullet$ & $\graybullet$ & $\graybullet$ & $\graybullet$ & $\graybullet$ \\
    \hspace{5pt}- Sub-node-level Slicing & $\graybullet$ & $\graybullet$ & $\graybullet$ & $\graybullet$ & $\graybullet$ & $\graybullet$ & $\graybullet$ & $\graybullet$ & & $\graybullet$ \\
    \hspace{5pt}- Automated Selection & $\graybullet$ & & & & & & & & & \\
    \midrule
    \textbf{Federation Support} & $\graybullet$ & & & & Unknown & & & & & $\graybullet$ \\
    \bottomrule
\end{tabularx}
\begin{tablenotes}
\setlength\itemsep{2pt} % let the notes breathe a bit more
\renewcommand{\UrlFont}{\rmfamily\tiny}
\item
\item \textbf{Short name} \tabto{2cm}\textbf{Name} \tabto{6.2cm}\textbf{First release} \tabto{8cm}\textbf{Source code} 
\item EdgeNet \tabto{2cm}EdgeNet \tabto{6.2cm}2019-10 \tabto{8cm}\url{https://github.com/EdgeNet-project/edgenet}
\item HNC \tabto{2cm}Hierarchical Namespace Controller \tabto{6.2cm}2019-11 \tabto{8cm}\url{https://github.com/kubernetes-sigs/hierarchical-namespaces}
\item Capsule \tabto{2cm}Clastix Labs' Capsule \tabto{6.2cm}2020-09 \tabto{8cm}\url{https://github.com/clastix/capsule}
\item kiosk \tabto{2cm}Loft's kiosk \tabto{6.2cm}2020-02 \tabto{8cm}\url{https://github.com/loft-sh/kiosk}
\item Arktos \tabto{2cm}Centaurus's Arktos \tabto{6.2cm}2020-04 \tabto{8cm}\url{https://github.com/CentaurusInfra/arktos}
\item VC \tabto{2cm}VirtualCluster \tabto{6.2cm}2021-06 \tabto{8cm}\url{https://github.com/kubernetes-sigs/cluster-api-provider-nested/tree/main/virtualcluster}
\item k3v \tabto{2cm}Rancher's k3v \tabto{6.2cm}2019-07 \tabto{8cm}\url{https://github.com/ibuildthecloud/k3v}
\item vcluster \tabto{2cm}Loft's vcluster \tabto{6.2cm}2021-04 \tabto{8cm}\url{https://github.com/loft-sh/vcluster}
\item Kamaji \tabto{2cm}Clastix Labs' Kamaji \tabto{6.2cm}2022-05 \tabto{8cm}\url{https://github.com/clastix/kamaji}
\item VK+ \tabto{2cm}\textbf{Virtual Kubelet based frameworks}
\item \tabto{2cm}Virtual Kubelet \tabto{6.2cm}2018-02 \tabto{8cm}\url{https://github.com/virtual-kubelet/virtual-kubelet}
\item \tabto{2cm}Liqo \tabto{6.2cm}2020-10 \tabto{8cm}\url{https://github.com/liqotech/liqo}
\item \tabto{2cm}Admiralty \tabto{6.2cm}2018-12 \tabto{8cm}\url{https://github.com/admiraltyio/admiralty} 
\item \tabto{2cm}Tencent's tensile-kube \tabto{6.2cm}2021-02 \tabto{8cm}\url{https://github.com/virtual-kubelet/tensile-kube}
\end{tablenotes}
\end{threeparttable}
\end{table*}

\renewcommand{\UrlFont}{\rmfamily\smaller}

\subsection{Multitenancy Approach}
\label{sec-multitenancy-approach}

The scientific literature describes two approaches to enabling CaaS multitenancy: multi-instance~\cite{guo2007framework}, and single-instance native~\cite{jia2021systematic}.
We ourselves further distinguish between multi-instance through multiple clusters and multi-instance through multiple control planes, making three approaches altogether, as shown in Table~\ref{tab:related-work-comparison}.
The approaches are illustrated in Fig.~\ref{fig:multitenancy-categories} and we describe them as follows:    

\begin{figure*}[t]
    \centering
    \subfloat[\textbf{Multi-instance through multiple clusters.} 
    Each tenant receives a separate cluster, including both the control plane and worker nodes. 
    This imposes considerable overhead.]{
        \includegraphics[width=0.31\linewidth]{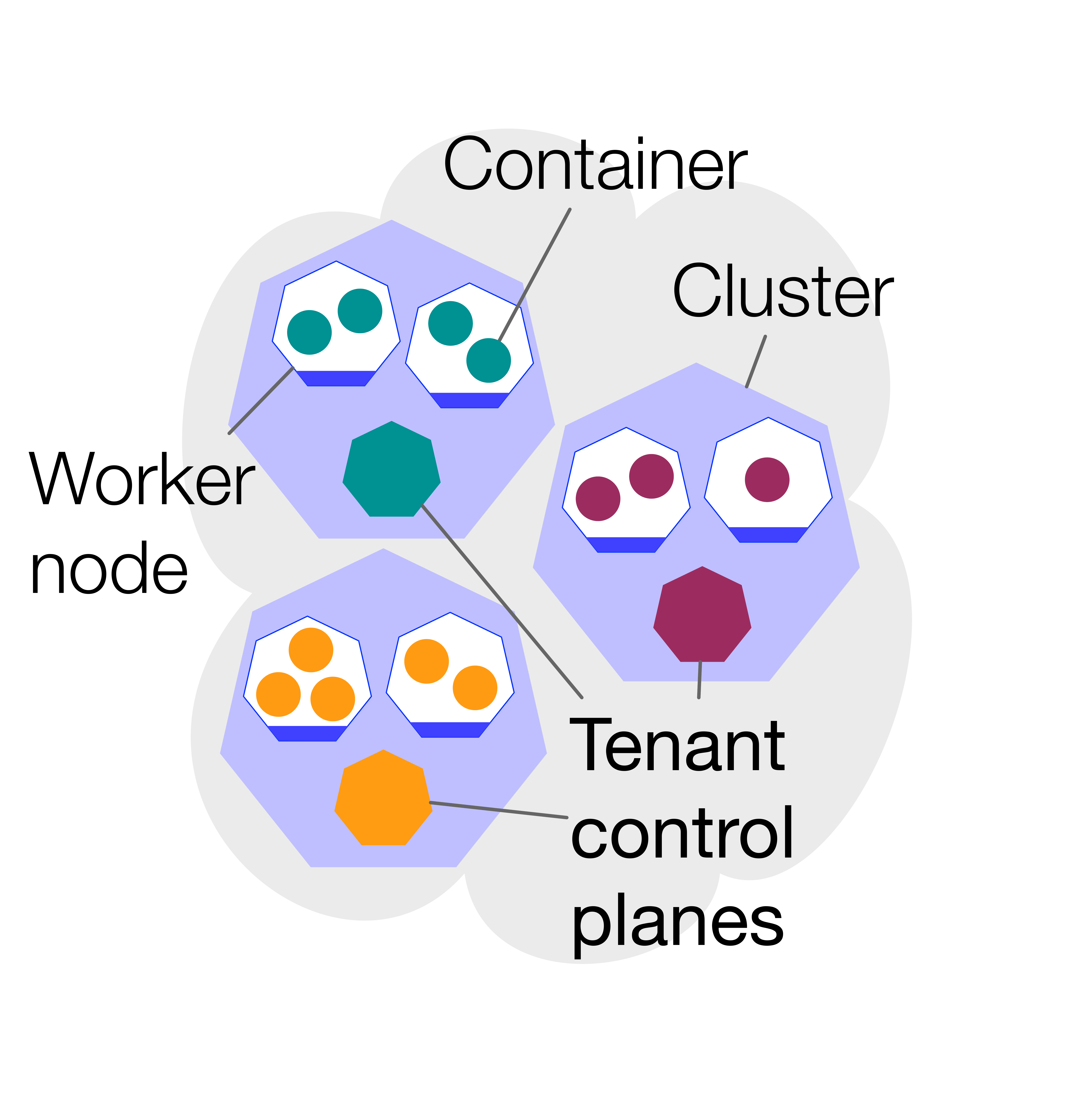}
        \label{fig:multi-instance-clusters}
    }
    \hfill
    \subfloat[\textbf{Multi-instance through multiple control planes.} 
    A physical cluster is divided into logical ones, each offered to a different tenant. 
    It can reuse worker nodes and the networking of the host cluster.]{
        \includegraphics[width=0.31\linewidth]{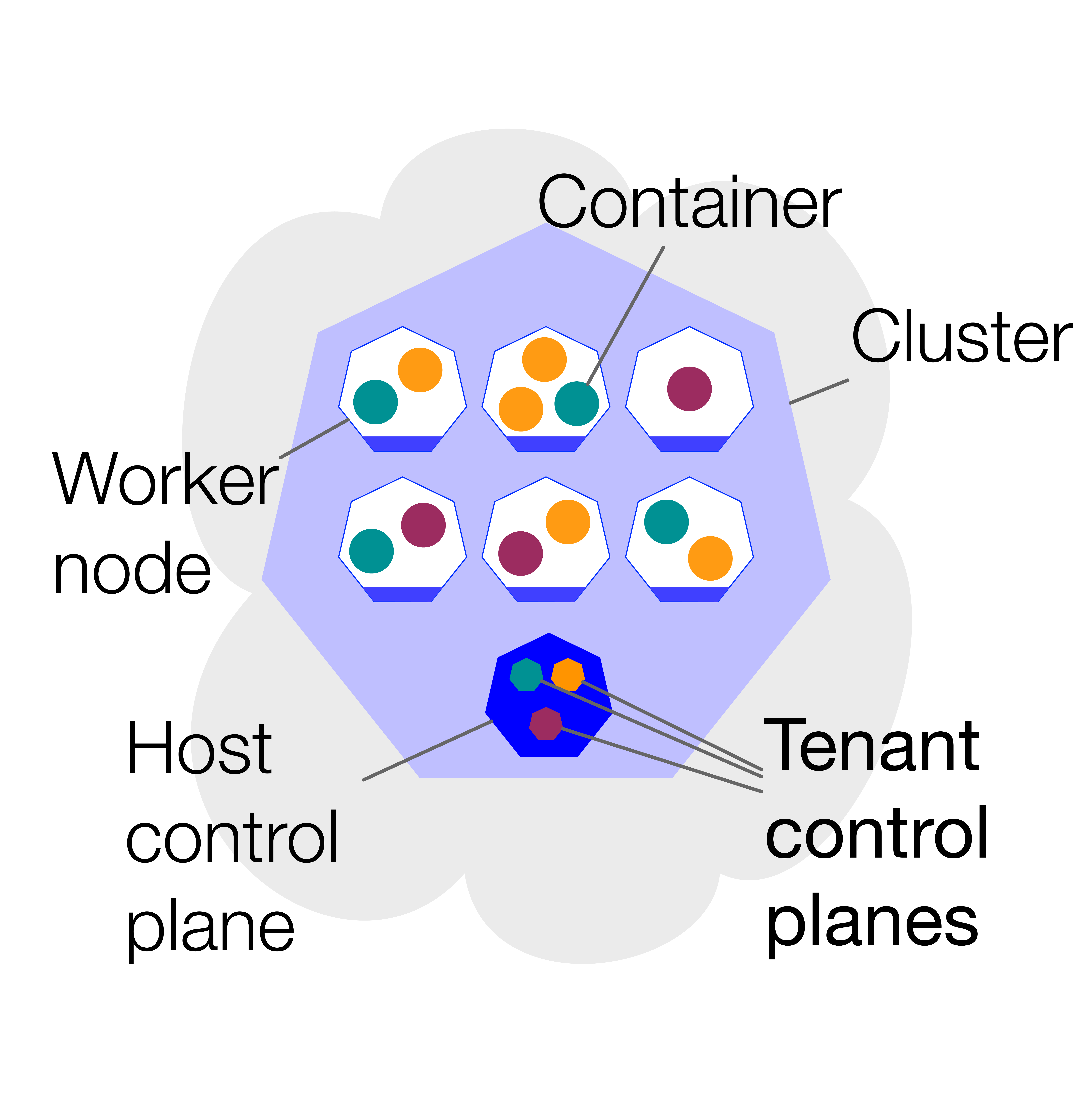}
        \label{fig:multi-instance-cps}
    }
    \hfill
    \subfloat[\textbf{Single-instance native.} 
    This low overhead approach has all tenants share a single cluster and a single control plane. Isolation between tenants is ensured by logical entities such as Kubernetes namespaces.]{
        \includegraphics[width=0.31\linewidth]{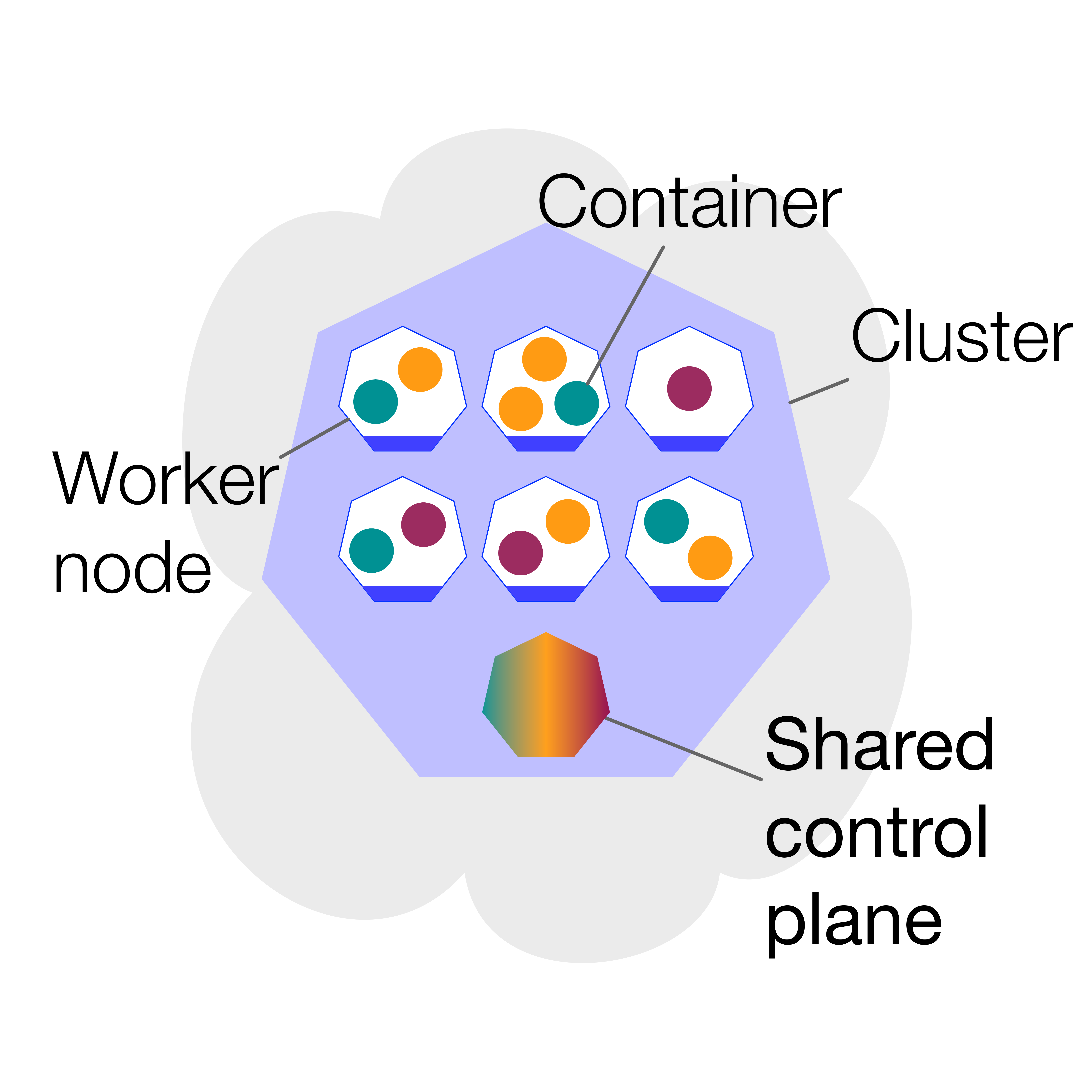}
        \label{fig:single-instance}
    }
    \caption{\textbf{Multitenancy Approaches.} 
    The multi-instance approaches provide each tenant with its own instance of the control plane (or, at the least, of certain control plane components) and, optionally, its own set of worker nodes, ensuring better isolation between tenants.
    The single-instance native approach caters to multiple tenants through a single control plane, while having them share %access to 
    the resources of a single set of worker nodes, thereby providing improved performance.}
    \label{fig:multitenancy-categories}
\end{figure*}

\subsubsection{Multi-instance through multiple clusters}

Fig.~\ref{fig:multi-instance-clusters} illustrates the multi-instance through multiple clusters approach, in which each tenant receives its own cluster.
The proprietary commercial CaaS offerings (see Table~\ref{tab:commercial-caas}) are structured in this way, but there is no open-source framework to enable precisely this form of multitenancy, spinning up and spinning down full Kubernetes clusters on demand for different tenants.
Existing open-source tools for deploying Kubernetes clusters, such as RKE\footnote{RKE \url{https://rke.docs.rancher.com/}} and Kubespray,\footnote{Kubespray \url{https://kubespray.io}} do not address multitenancy.

There is, however, a set of open-source Kubernetes frameworks that do address multitenancy for the case in which there are already multiple tenants, each of which possesses one or more of their own clusters, even if these frameworks do not spin up or spin down the clusters on demand.
These frameworks, based on the code of \textbf{Virtual Kubelet}~\cite{virtualkubelet2020virtualkubelet}, a sandbox project of the Cloud Native Computing Foundation, are designed to allow workloads from one cluster to be deployed to another cluster.
Their primary focus is on cross-cluster deployment in general, and multitenancy arises only in the specific case of clusters belonging to different tenants, but since they do enable this sort of multitenancy, we examine the advantages and disadvantages of doing so.

\begin{figure}
    \centering
    \includegraphics[width=0.31\columnwidth]{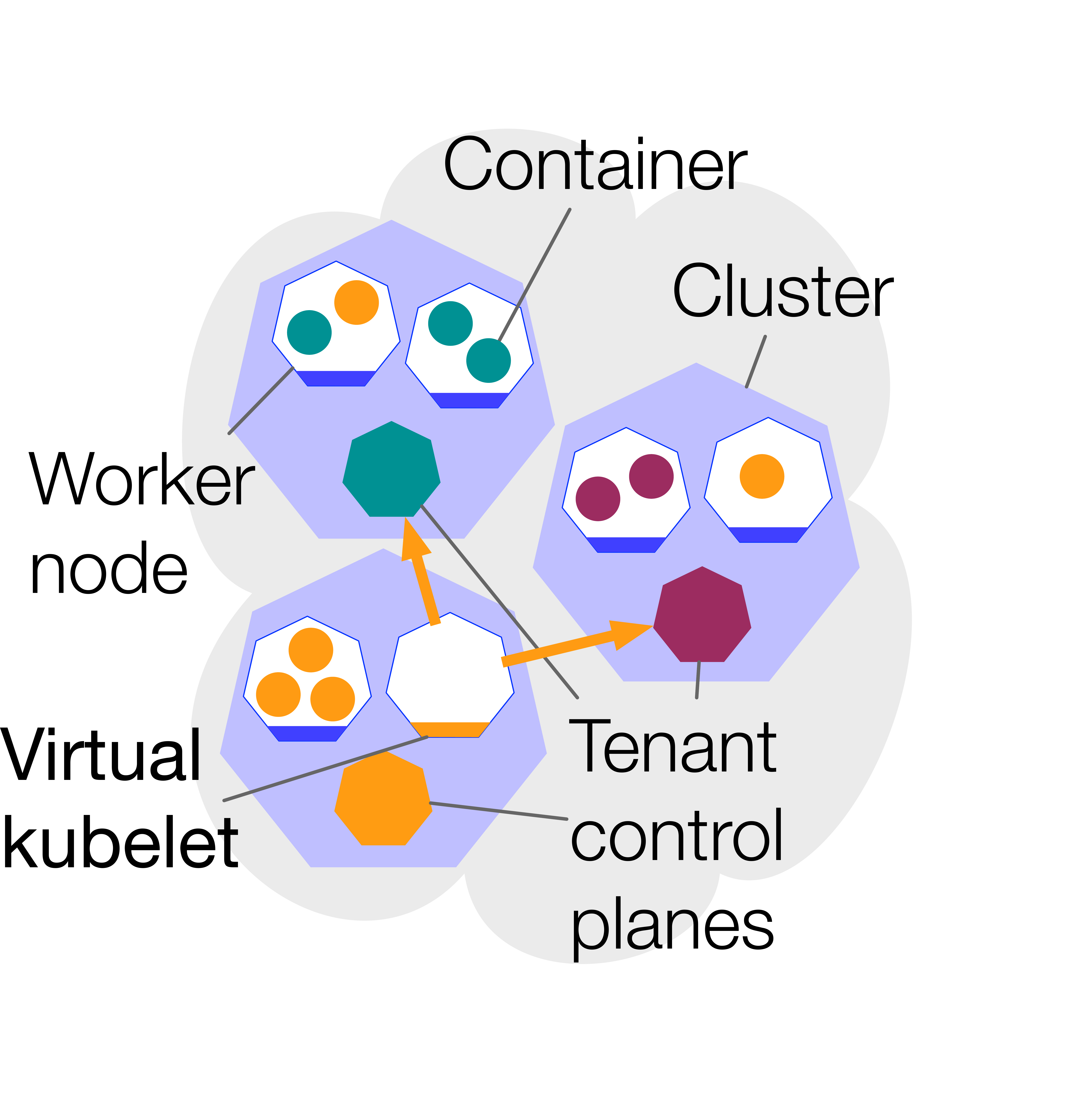}
    \caption{\textbf{Multitenancy through Virtual Kubelet.} The virtual kubelet masquerades as the kubelet of a node in a cluster, but in reality deploys workloads from that cluster to other clusters via those clusters' APIs. When a local cluster belongs to one tenant and a remote cluster belongs to another tenant, this results in a distinctive form of multitenancy, with clusters that, while otherwise belonging to one tenant, host pods from other tenants.}
    \label{fig:virtual-kubelet}
\end{figure}

As illustrated in Fig.~\ref{fig:virtual-kubelet}, Virtual Kubelet establishes a connection from one cluster to another by leveraging Kubernetes' \textit{kubelet}\footnote{Kubernetes documentation: \textit{kubelet} \url{https://kubernetes.io/docs/concepts/overview/components/\#kubelet}} API.  
A kubelet is the agent that runs on each node of a Kubernetes cluster in order to manage the life cycles of pods, which are groups of containers associated with a workload.
By implementing the kubelet API, a virtual kubelet masquerades as the kubelet of an individual node, but is in reality a stand-in for the remote cluster.
It, in turn, uses the remote cluster's control plane API to deploy and manage workloads on that cluster.

Although we might think of this as a small scale form of federation, the Virtual Kubelet authors expressly say that it ``is not intended to be an alternative to Kubernetes federation,'' by which we understand a full-featured and scalable federation.
Similarly, as we have mentioned, Virtual Kubelet is not primarily designed for multitenancy.
By contrast, EdgeNet is designed precisely for federation and multitenancy.
While similar to Virtual Kubelet in the sense that EdgeNet introduces agents to transfer workloads from one cluster to another, EdgeNet avoids the overhead associated with each tenant having its own cluster.
This is because, in EdgeNet, it is the cloud and edge cloud providers that possess the clusters.
Provider ownership of the clusters also means that an EdgeNet tenant can rely upon a provider to ensure the privacy of its workloads, rather than relying upon another tenant to do so.

\textbf{Liqo}~\cite{liqo2020liqo}, \textbf{Admiralty}~\cite{admiralty2018admiralty} and \textbf{tensile-kube}~\cite{tencent2021tensile} are all based on the Virtual Kubelet code.
Liqo is one of the few frameworks to date to be the subject of a peer-reviewed scientific paper~\cite{iorio2022liqo}.
The authors are careful to state that some of the issues that arise from multitenancy, such as the manner in which the workloads of different tenants in the same cluster are isolated from each other, remain to be addressed.\footnote{From the Liqo paper~\cite{iorio2022liqo}: ``Specifically, we foresee a \textit{shared security responsibility model}, with the provider responsible for the creation of well-defined sandboxes and the possible provisioning of additional security mechanisms (e.g., secure storage) negotiated at peering time.''}
Sec.~\ref{sec-container-isolation} describes our proposed resolution for this problem.

\subsubsection{Multi-instance through multiple control planes}

In the multi-instance through multiple control planes approach, all tenants are supported by a single cluster, but each tenant acquires its own control plane within that cluster, as illustrated by Fig.~\ref{fig:multi-instance-cps}.
One or more nodes are dedicated to supporting the tenant control planes, and, within each control plane node, containers, or containers grouped into pods, isolate one tenant's control plane from another's.
(Isolating control planes from each other through containers imposes lower overhead than doing so with VMs.)
There are variants to this approach, in which some control plane components, like the scheduler, are shared among tenants, while others, such as the API server and database, are duplicated so as to provide one instance to each tenant.
In any case, this approach gives each tenant a full view of its own control plane view, which it can use for customizing its own environment.
            
Frameworks that follow this approach differ in how they isolate tenant workloads from each other.
If tenants share a common set of worker nodes as they do in VirtualCluster, k3v, and vcluster, the degree of isolation will depend upon the container runtime used to run the containers. 
If each tenant acquires its own dedicated set of worker nodes, as happens in Kamaji, then there is better isolation.

\textbf{VirtualCluster}~\cite{virtualcluster2021virtualcluster} 
% (abbreviated to VC in Table~\ref{tab:related-work-comparison}, and not to be confused with vcluster) 
is one of the two open-source frameworks incubated by the Kubernetes Multi-Tenancy Working Group.
It virtualizes the control plane components per tenant, with the exception of the scheduler.
For isolation between the worker nodes of different tenants, it uses Kata containers~\cite{randazzo2019kata}.

A drawback of VirtualCluster is the cost of providing separate control plane components per tenant.
In a peer-reviewed scientific paper~\cite{zheng2021multi}, the VirtualCluster authors state that this cost is a blocking point when more than a thousand tenants are in the cluster. 
By contrast, EdgeNet's shared control plane approach allows far more tenants to be allowed into a given cluster, and allows more tenants to arrive within a short period of time, as we show in Sec.~\ref{sec-benchmarking}.
In a federated edge cloud environment, where we anticipate limited resources, large numbers of workloads, and the rapid propagation of workloads from one cluster to another, the shared control plane approach has a clear advantage.
In fairness to VirtualCluster, it is designed for a different sort of environment.

In Rancher's \textbf{k3v}~\cite{rancher2019k3v}, the control plane is virtualized on a per-tenant basis, similar to VirtualCluster, but it does not provide data plane isolation, as VirtualCluster does.
Exceptionally among the frameworks, k3v does not provide a mechanism for managing tenant resource quotas, as we mention in Sec.~\ref{sec-tenant-resource-quota}.
 
\textbf{vcluster}~\cite{loft2021vcluster}, not to be confused with VirtualCluster, is one of two open-source frameworks developed by Loft, the other being kiosk, which takes the single-instance native approach.
In the control plane, each vcluster has a separate API server and data store. 
Workloads created on a vcluster are copied into the namespace of the underlying cluster to be deployed by the shared scheduler. 

\textbf{Kamaji}~\cite{clastix2022kamaji} is one of two open-source frameworks developed by Clastix Labs, the other being Capsule, which takes the single-instance native approach.
Kamaji enables Kubernetes multitenancy by running tenant control planes as pods on a common cluster, known as the admin cluster.
Each tenant receives its own dedicated worker nodes.
Isolation between worker nodes on the same machine is enabled through VMs, which introduces much higher overhead than would isolation through containers.

\subsubsection{Single-instance native}

In the single-instance native approach, all tenants share a single control plane and a common set of worker nodes, as illustrated in Fig.~\ref{fig:single-instance}.
Control plane isolation is ensured through a logical entity, such as Kubernetes namespaces, that introduces negligible overhead, but provides less control plane isolation compared to a multi-instance approach.
Workload isolation depends upon the container runtime, as it does for the multi-instance through multiple clusters that are federated approach and for the multi-instance through multiple control planes approach.
    
This approach demands significant coding work to give each tenant an experience akin to using their own separate cluster.

The single-instance native approach's scaling advantage is illustrated by a scenario examined by Guo et al.\@ in which it supported thousands of tenants, as opposed to just dozens for a multi-instance approach~\cite{guo2007framework}.
It also has lower operational costs~\cite{chong2006multi}.
And it is lighter weight for workload mobility, allowing containers to be spun up and spun down with less overhead than in a multi-instance approach, as we show through benchmarking in Sec.~\ref{sec-benchmarking}.
For these reasons, we have adopted the single-instance native approach for EdgeNet.

The Hierarchical Namespace Controller (\textbf{HNC}) is one of the two open-source frameworks incubated by the Kubernetes Multi-Tenancy Working Group, the other being VirtualCluster.
HNC takes the single-instance native approach, whereas VirtualCluster takes the multi-instance through multiple control planes approach.
HNC uses a hierarchical namespace structure in order to enable multitenancy.\footnote{Kubernetes Multi-tenancy Working Group documentation: \textit{HNC: Concepts} \url{https://github.com/kubernetes-sigs/hierarchical-namespaces/blob/master/docs/user-guide/concepts.md}}
Functionalities such as policy inheritance that allow objects to be replicated across namespaces are built upon this hierarchy.

Aspects of this work that have inspired our own multitenancy framework are its hierarchical namespace structure and the terminology that it employs.
We have also designed our own framework to avoid what we perceive to be its defects:
\begin{itemize}
    \item HNC does not enforce unique names for namespaces, opening the possibility for namespace conflicts.
    \item HNC's quota management system is not aligned with the hierarchical namespace structure so as to limit a child's quota based upon its parent's quota, though community documentation states\footnote{Kubernetes Multi-tenancy Working Group documentation: \textit{HNC: Policy inheritance and object propagation} \url{https://github.com/kubernetes-sigs/hierarchical-namespaces/blob/master/docs/user-guide/concepts.md\#policy-inheritance-and-object-propagation}} that work is underway to enable this.
    \item HNC's quota management system allows namespaces without quota to coexist alongside namespaces that have quotas, which puts those quotas at risk (see Fig.~\ref{fig:resource-quota-hnc} and discussion in Sec.~\ref{sec-tenant-resource-quota}).
\end{itemize}

\textbf{Capsule}~\cite{clastix2021capsule} is one of two open-source frameworks developed by Clastix Labs, the other being Kamaji, which, as we have seen, takes the multi-instance through multiple control planes approach.
Capsule is one of two frameworks that adopts flat namespaces (see Sec.~\ref{sec-tenant-environment-customization}) as its customization approach, the other being kiosk.
It gives a tenant the possibility of creating resources that can be replicated across a collection of namespaces of the tenant, and it provides the cluster administrator with the possibility to copy resources among namespaces of various tenants.
Although this approach facilitates the management of multiple namespaces that belong to a tenant, so it eases management complexity, it may not be fully scalable for extensive tenant settings, as we discuss in the following subsection.
Capsule aims at allowing an organization to share a single cluster efficiently, hence not accounting for the needs of the envisaged edge computing infrastructure.
 
\textbf{kiosk}~\cite{loft2020kiosk} is one of two open-source frameworks developed by Loft, the other being vcluster, which vcluster takes the multi-instance through multiple control planes approach.
This solution uses flat namespaces approach, as does Capsule, for customization.
A tenant is represented by an abstraction called an \emph{account}, and an account can create a namespace through an entity called a \emph{space}.
Each space is strictly tied to only one namespace.
This framework permits the preparation of templates that can be employed during namespace creation, facilitating the automated provisioning of resources as defined within these templates in the designated namespaces.
Despite alleviating management complexity, this approach still shares Capsule's limitations stemming from flat namespaces.
Multi-cluster tenant management is listed on their roadmap, but the project does not seem to be under active development, as the latest commit in its main branch was around a year ago.
 
Centaurus's \textbf{Arktos}~\cite{centaurus2020arktos} takes the single-instance native approach to multitenancy.
As discussed in Sec.~\ref{sec-tenant-environment-customization}, it is the only framework that takes a tenant-wise abstraction approach to enabling customization.
Arktos achieves this through API modifications,\footnote{Arktos documentation: \emph{Multi-tenancy Overview} \url{https://github.com/CentaurusInfra/arktos/blob/master/docs/design-proposals/multi-tenancy/multi-tenancy-overview.md\#api-server}} which may require a significant amount of effort to keep aligned with the upstream Kubernetes control plane code.
Its architecture primarily consists of three main software entities: an \emph{API gateway} that receives tenant requests, a \emph{Tenant Partition (TP)} that gives the illusion of each tenant acquiring an individual cluster, and a \emph{Resource Partition (RP)} that operates on resources like nodes~\cite{dubey2022usermanagement}. 
Although not all of its features are precisely presented, based upon our reading of their documentation, we consider that this solution addresses some federation aspects, such as scalability and cloud-edge communication.
They provide a vision of consolidating 300,000 nodes belonging to different resource partitions into a single regional control plane.
However, the main branch of their project repository has not received commits for around a year, implying that it may not be currently undergoing active development.

\subsection{Customization Approach}
\label{sec-tenant-environment-customization}
Containers-as-a-service cannot scale to a large number of tenants if the mechanism by which each tenant obtains the environments in which to deploy its workloads, and configures each environment to meet the needs of its workload, requires manual intervention at every stage by the cloud administrator.
Each tenant should have a degree of autonomy to: create and delete the environments in which its workloads can be deployed;
obtain resource quotas and assign them to those environments;
and designate users for the environments, assign roles to those users, and grant permissions based upon those roles.
Some combination of automation of these processes and delegation of administrative responsibility is needed to enable that autonomy.
In Table~\ref{tab:related-work-comparison}, we call the way in which a multitenancy framework does this its \textit{Customization Approach}.

By giving each tenant its own control plane, which the tenant's administrator can use to configure its environments as they wish, the multi-instance frameworks provide the greatest flexibility.
We call this approach the \textbf{Full Control Plane View}.
As Table~\ref{tab:related-work-comparison} shows, it is offered by the frameworks that follow the multi-instance through multiple clusters approach (Virtual Kubelet based frameworks), since each cluster has its own control plane, and, of course, by the multi-instance through multiple control planes approach (VirtualCluster, k3v, vcluster, and Kamaji).

Some of these frameworks (Kamaji and, partially, Virtual Kubelet based frameworks) allow additional server environment configuration to take place by enabling SSH access to worker nodes, and this is noted as \textbf{Data Plane} customization in the comparison table.
In Virtual Kubelet based frameworks, administrators of a tenant that owns a cluster can typically access the worker nodes in that cluster by SSH, but not the ones in other clusters, and this is classified as \textbf{Partial} in the comparison table.

In frameworks that follow the single-instance native multitenancy approach, some extensions to Kubernetes are required in order to safely enable customization.
This is because in standard Kubernetes, giving a tenant's administrator the permissions necessary to configure their own environments means giving them the ability to configure other tenants' environments as well.
Since there is no control plane isolation mechanism other than namespaces, an administrator who has permission to create, modify, and delete namespaces can do so freely across the board.
Rather than hand out such permissions, a single-instance customization approach needs to provide one or more custom resources that a tenant's administrator can access, and the controllers of those will ensure safety while configuring the tenant environment on the administrator's behalf.

Among the single-instance frameworks, Arktos employs the most elaborate customization approach: that of introducing a new abstraction, beyond namespaces, by which to isolate tenants from one another in the control plane.
As this abstraction is meant to capture the notion of a tenant, we refer to it in Table~\ref{tab:related-work-comparison} as a \textbf{Tenant-wise Abstraction}.
Our concern about this approach is the amount of development work that it might entail, both to develop this new abstraction and to maintain its compatibility with Kubernetes' upstream version of the control plane code.

Instead of introducing an entirely new abstraction, frameworks can build on Kubernetes' existing control plane isolation mechanism: namespaces.
We identify two ways of doing so.
The simpler one, followed by Capsule and kiosk, is to follow the standard Kubernetes approach, in which each namespace exists independently of every other namespace.
This is described as \textbf{Flat Namespaces} in Table~\ref{tab:related-work-comparison}.

Another way, but one that requires more development work, is to provide controllers that keep track of the relationships between namespaces, such as several namespaces all belonging to the same tenant.
Since the two frameworks that do this, EdgeNet and HNC, do so by maintaining a hierarchical structure through which to track the relationships, we identify this approach as \textbf{Hierarchical Namespaces} in Table~\ref{tab:related-work-comparison}.

\begin{figure*}
    \centering
    \includegraphics[width=0.8\linewidth]{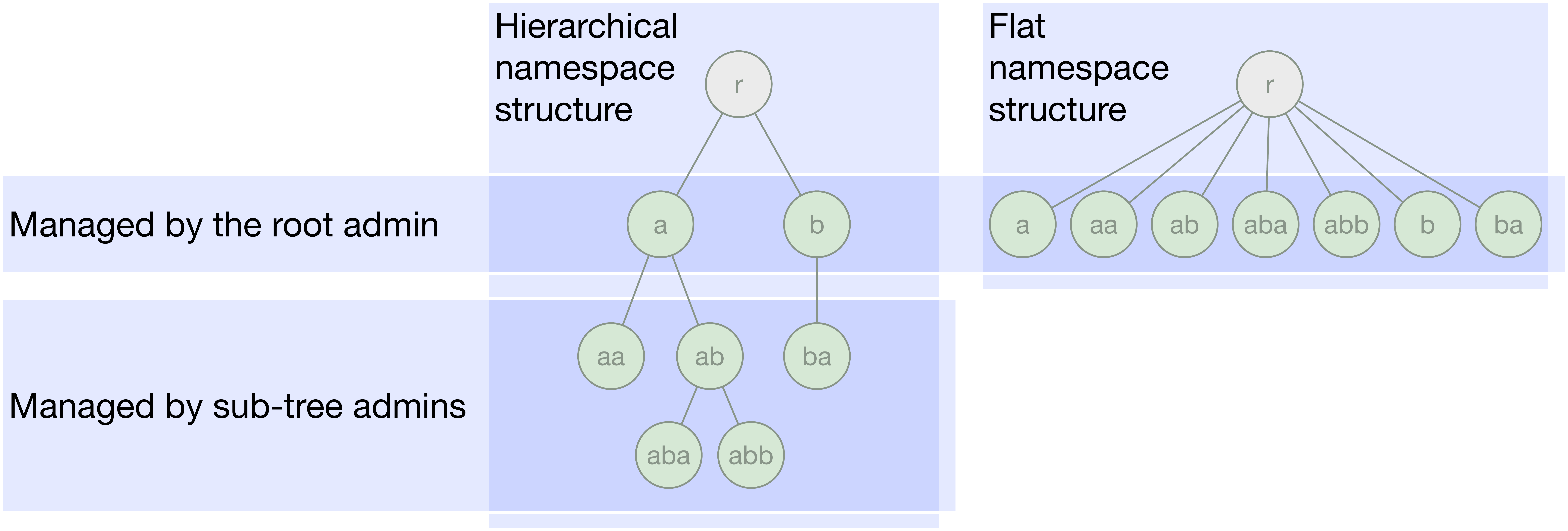}
    \caption{\textbf{Customization Approach: Hierarchical versus Flat Namespaces.} 
    The same seven namespaces organized into a hierarchy (left) and without a hierarchy (right), in each case under a root environment \textit{r}, which is not itself a namespace. 
    \newline \newline The hierarchy captures relationships between the namespaces: \textit{a} and \textit{b} are the core namespaces belonging to two tenants, whereas the others belong to sub-trees of those core namespaces. 
    For example, \textit{aa} and \textit{ab} are subnamespaces of \textit{a}. They belong to the same tenant as \textit{a} and they may inherit a portion of that tenant's resource quota, %users, 
    user roles, and the permissions that accompany those roles. 
    Likewise, \textit{aba} and \textit{abb} belongs to the same tenant as \textit{ab} and may inherit from it. 
    Management of tasks such as the approval of new namespaces, and the modification of quotas, users, etc., can be delegated to each tenant's administrators, and, further down the hierarchy, to sub-tree administrators.
    \newline \newline The flat structure does not express these relationships.
    For example, no mechanism provides for \textit{aa} to inherit from \textit{a}.
    If they are to share configuration parameters, this needs to be expressly requested by the common administrator of the two namespaces.
    There are efforts to solve this issue through configuration templates to be applied to multiple namespaces. 
    Nevertheless, as the number of namespaces that a tenant has grows, it results in management complexity for the root admin of this tenant, which makes it challenging to keep track of independent namespaces.
    }
    \label{fig:namespaces_hierarchical_v_flat}
\end{figure*}

Fig.~\ref{fig:namespaces_hierarchical_v_flat} compares the two namespace structures.
A hierarchical structure permits configurations to be inherited and allows for configuration tasks to be delegated, offloading tasks from administrators at the top of the hierarchy to administrators further down.
The prime disadvantage of a flat namespace structure is that, even with automation, the root admins of tenants are highly solicited.
EdgeNet adopts a hierarchical namespace structure, which is implemented by the architecture described in Sec.~\ref{sec-architecture-tenant} and Sec.~\ref{sec-architecture-subnamespace}.

\subsection{Consumer and vendor modes}
\label{sec-consumer-and-vendor-modes}

Cloud services generally support two types of tenancy: \textbf{Consumer Mode}, in which the tenant is the end user of the resources; and \textbf{Vendor Mode}, in which the tenant can resell access to the resources to others.

The type of tenancy affects the visibility that the manager of a tenant has into that tenant’s isolated environments.
For a consumer tenant, these environments are generally termed \textit{workspaces}, and they are created to be used by the members of that tenant’s group or organization. 
A manager of a set of workspaces needs visibility into who the users of each workspace are, and needs fine-grained control over the rights of those users with respect to those workspaces. 
But a vendor tenant manages a set of subtenant environments that are destined for its own customers.
A customer expects a certain level of privacy, with the users and user rights of their subtenant environment remaining hidden from the vendor.

As shown in Table~\ref{tab:related-work-comparison}, all of the CaaS multitenancy frameworks that we have studied support consumer tenancy, but only EdgeNet and Virtual Kubelet based frameworks support vendor tenancy.
We expect that the same commercial logic that has driven other cloud service models towards both forms of tenancy will lead to support for vendor tenancy being generalized for containers-as-a-service.

In order to enable any sort of tenancy, a system must support authorization and isolation mechanisms.
It requires greater expressiveness to support both consumer and vendor tenancy than it does to support consumer tenancy alone.
Such expressiveness, for example, allows a tenant to create a subtenant for the purpose of reselling its own allocated resources.
This can be done in different ways depending upon the multitenancy approach:
\begin{itemize}
    \item \emph{Multi-instance through multiple clusters:} A tenant who owns a cluster can open this cluster for use by one of its subtenants.
    Because of the ease of doing so, we indicate Virtual Kubelet based frameworks as offering support for a vendor mode, even though their documentation does not explicitly mention this.
    However, since such an approach requires a cluster per tenant, this introduces high overhead, as our benchmarking shows in Sec.~\ref{sec-benchmarking}.
    \item \emph{Multi-instance through multiple control planes:} A tenant could create a subtenant generated with its subtenant control plane instance running on top of the tenant control plane instance. 
    None of the frameworks that we have studied currently do this.
    \item \emph{Single-instance native:} A tenant can create a subtenant assigned with private namespaces that the tenant is solely authorized to remove.
    EdgeNet, having adopted the single-instance native approach to multitenancy, builds consumer and vendor modes on top of its hierarchical namespace structure.
    The implementation is described in Sec.~\ref{sec-architecture-subnamespace-modes} and illustrated in Fig.~\ref{fig:consumer_and_vendor_EdgeNet}.
\end{itemize}

\subsection{Tenant resource quota allocation}
\label{sec-tenant-resource-quota}

Resource quotas are popular in commercial settings, where they provide a basis for providers to bill their customers.
In situations where resources are constrained, quotas are also a simple means by which to ensure an equitable allocation of those resources.
Quotas are commonly used in the cloud, and Kubernetes supports them by providing a mechanism for allocating quotas to namespaces.\footnote{Kubernetes documentation: \textit{Resource Quotas} \url{https://kubernetes.io/docs/concepts/policy/resource-quotas/}}
The Kubernetes mechanism is conceived for the relatively small scale scenario of a single organization using a cluster, and an administrator who manually sets resource quotas per namespace so as to share out the resources among different teams in the organization. 
A multitenancy framework that is built on Kubernetes needs to automate this process, to enable it to scale.

As Table~\ref{tab:related-work-comparison} shows, all of the Kubernetes multitenancy frameworks that we have studied offer a mechanism for managing tenant resource quotas, with the exception of k3v.
We classify k3v in this way as we consider its mechanism to be incomplete.
In that framework, which is no longer under active development, a cluster administrator can set a resource quota in the host namespace of a virtual cluster, but the tenant will not be aware of it.

In the edge cloud, we can expect resources to be more constrained than in the cloud, and so the need for a quota allocation mechanism is even stronger.
Since our EdgeNet framework is designed for the edge cloud as well as the cloud, such a mechanism is a required feature of the framework.

\begin{figure*}[t]
    \centering
    \subfloat[\textbf{EdgeNet example.} The quota allocated to a sub-tree must be divided among a portion reserved for the namespace at which this sub-tree is rooted and the portions allocated to each subnamespaces.
    ]{
        \includegraphics[width=0.42\linewidth]{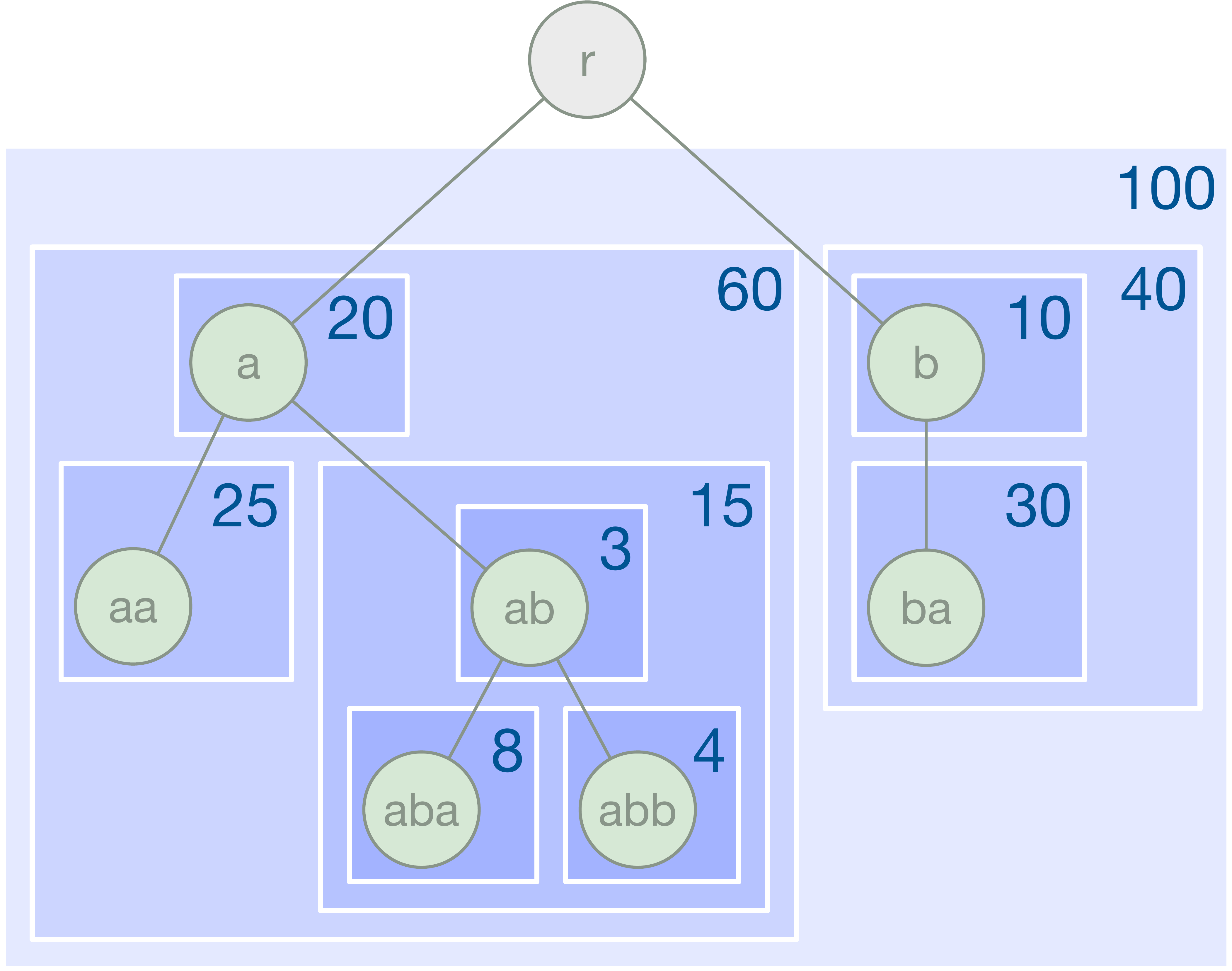}
        \label{fig:resource-quota-edgenet}
    }
    \hspace{0.1\linewidth}
    \subfloat[\textbf{HNC example.} In the same hierarchy as sub-trees that are constrained by quotas, it is possible to have sub-trees that are not constrained in this way.
    ]{
        \includegraphics[width=0.42\linewidth]{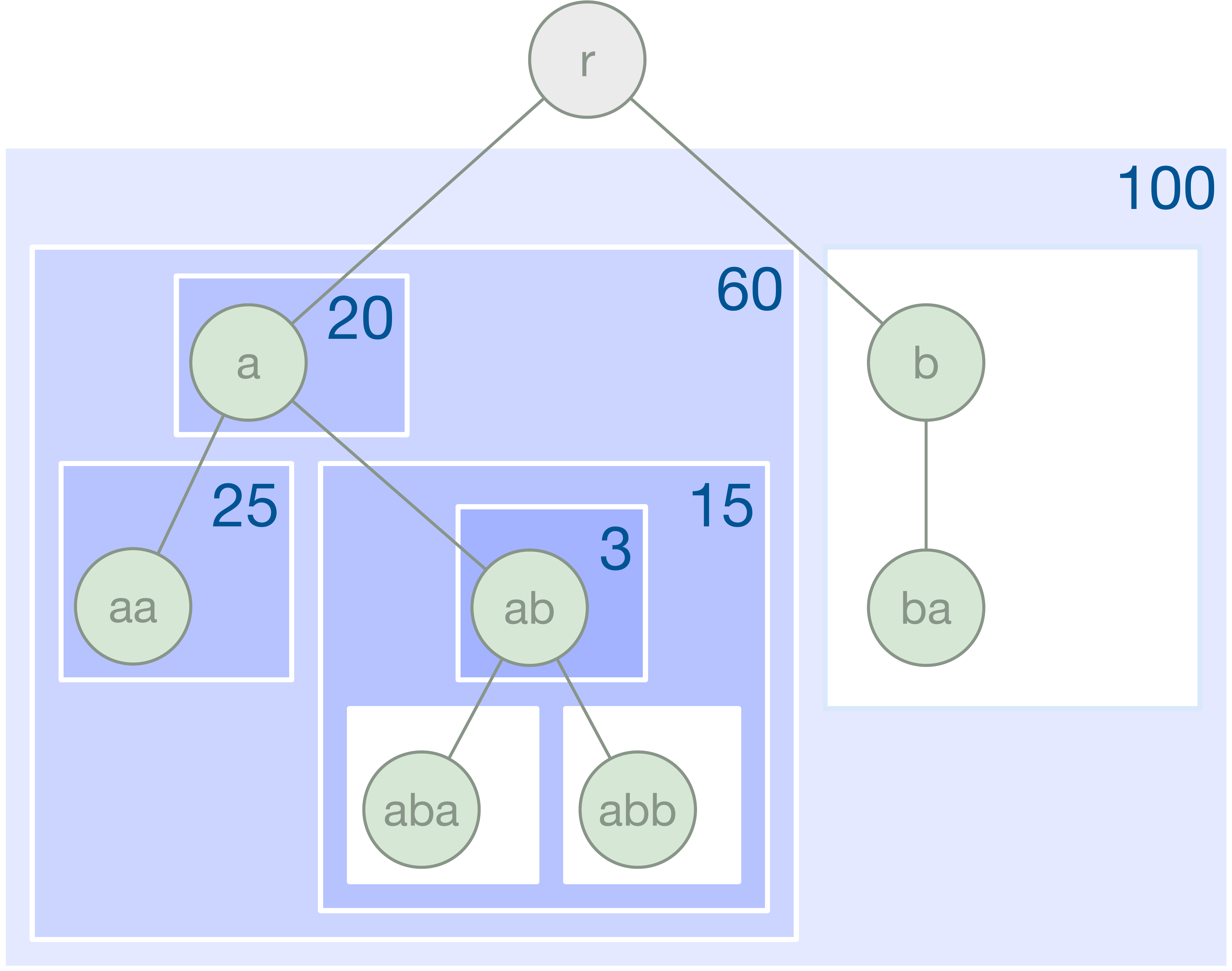}
        \label{fig:resource-quota-hnc}
    }
    \vspace{12pt}
    \caption{\textbf{Hierarchical allocation of resource quotas.} 
    Examples of a quota of $100$ being divided up among the sub-trees of a hierarchical namespace rooted at \textit{r}.
    The tenant of the sub-tree rooted at \textit{a} has been allocated a quota of $60$, from which it reserves $20$ for its core namespace and allocates $25$ and $15$ to the sub-trees rooted at \textit{aa} and \textit{ab}, respectively.
    \newline \newline In EdgeNet, the quota of $15$ must also be distributed within the sub-tree rooted at \textit{ab}.
    For example, here, $3$ is reserved for %that core namespace
    the namespace \textit{ab} and $8$ and $4$ are allotted to the sub-trees rooted at \textit{aba} and \textit{abb}, respectively.
    Likewise, quota must be allocated to the sub-tree rooted at \textit{b} and distributed within that sub-tree.
    \newline \newline
    HNC, on the other hand, allows portions of the hierarchy to be free of quotas.
    In this example, in HNC, the administrator of namespace \textit{ab} has, perhaps inadvertently, not set quotas for its subnamespaces, and likewise for the tenant administrator of \textit{b}.
    If workloads in \textit{aba} and \textit{abb} were to exceed a resource consumption of $12$ or the workloads at $b$ were to consume resources exceeding $40$, other namespaces with quotas might not be able to fully enjoy the resources quotas that had been reserved for them.}
    \label{fig:resource-quota-comparison}
\end{figure*}

Having made the design decision to use a hierarchical namespace structure, our quota management system needs to follow that structure.
This means building in dependencies between quotas: as shown in Fig.~\ref{fig:resource-quota-edgenet}, at each node in the namespace tree, quota must be shared out between the parent namespace located at that node and the sub-trees that are rooted at the children of that node.
EdgeNet's quota implementation is more thoroughly described in Sec.~\ref{sec-architecture-trq}.

The only other framework that uses hierarchical namespaces, HNC, also allows quota to be shared out hierarchically.
The mechanism employed in doing so relies on Google Cloud's Hierarchy Controller\footnote{Google Cloud Anthos: Hierarchical Resource Quotas  \url{https://cloud.google.com/anthos-config-management/docs/how-to/using-hierarchical-resource-quotas}} as its foundation.
But since it does not require that a quota be attributed to each namespace, it can end up constraining some namespaces while not constraining others, opening the possibility for a sub-tree to not enjoy the full resource quota that it has been allocated, as shown in Fig.~\ref{fig:resource-quota-hnc}.
In EdgeNet, quotas apply either to the entire tenant namespace hierarchy or not at all, so this problem cannot arise.

Resource quotas can be wasteful of resources if they are not used fully, while best-effort distribution of resources is more efficient without providing guarantees.
None of the Kubernetes multitenancy frameworks provides an intermediate solution.
Providing such a solution is on the EdgeNet development road map. 

\subsection{Variable slice granularity}
\label{sec-variable-slice-granularity}

\begin{figure*}
    \centering
    \subfloat[\textbf{Node-level slicing of servers.} An entire node that is a physical server is made available to a tenant.]{
        \includegraphics[width=0.42\linewidth]{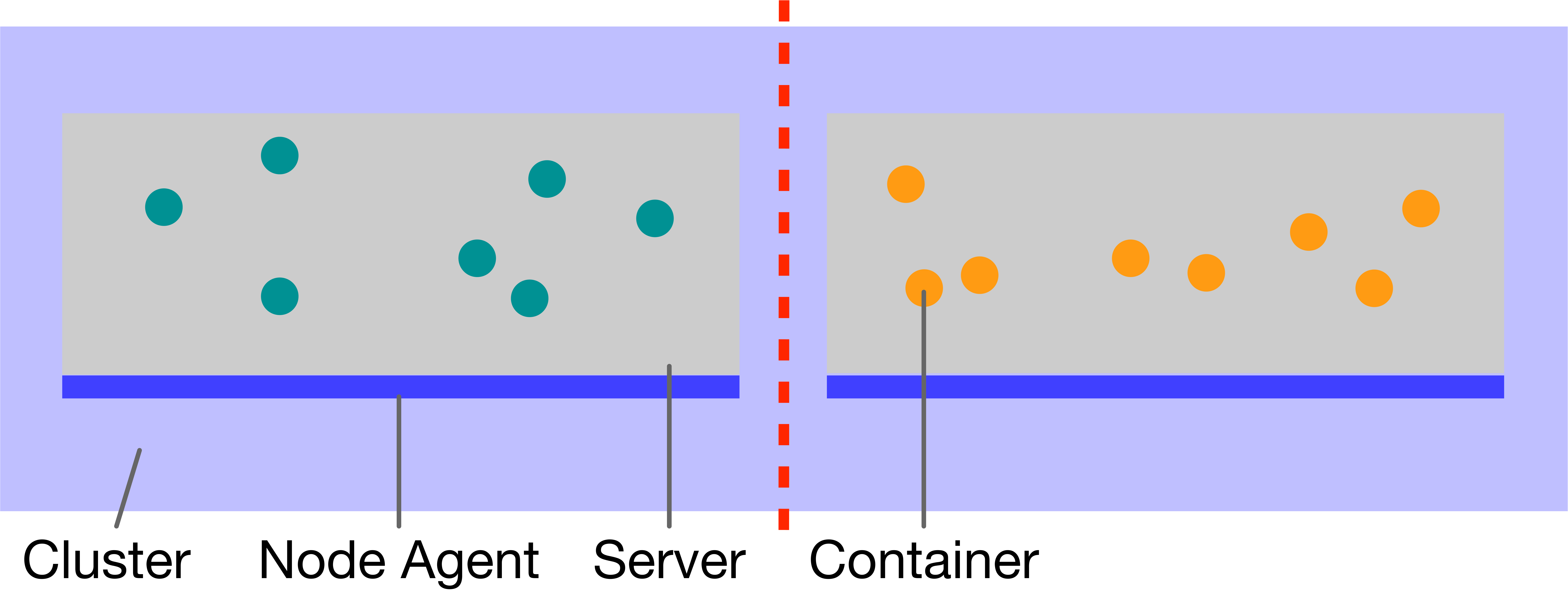}
        \label{fig:slicing-servers-node-level}
    }
    \hspace{0.1\textwidth}
    \subfloat[\textbf{Node-level slicing of VMs.} An entire node that is a VM is made available to a tenant.]{
        \includegraphics[width=0.42\linewidth]{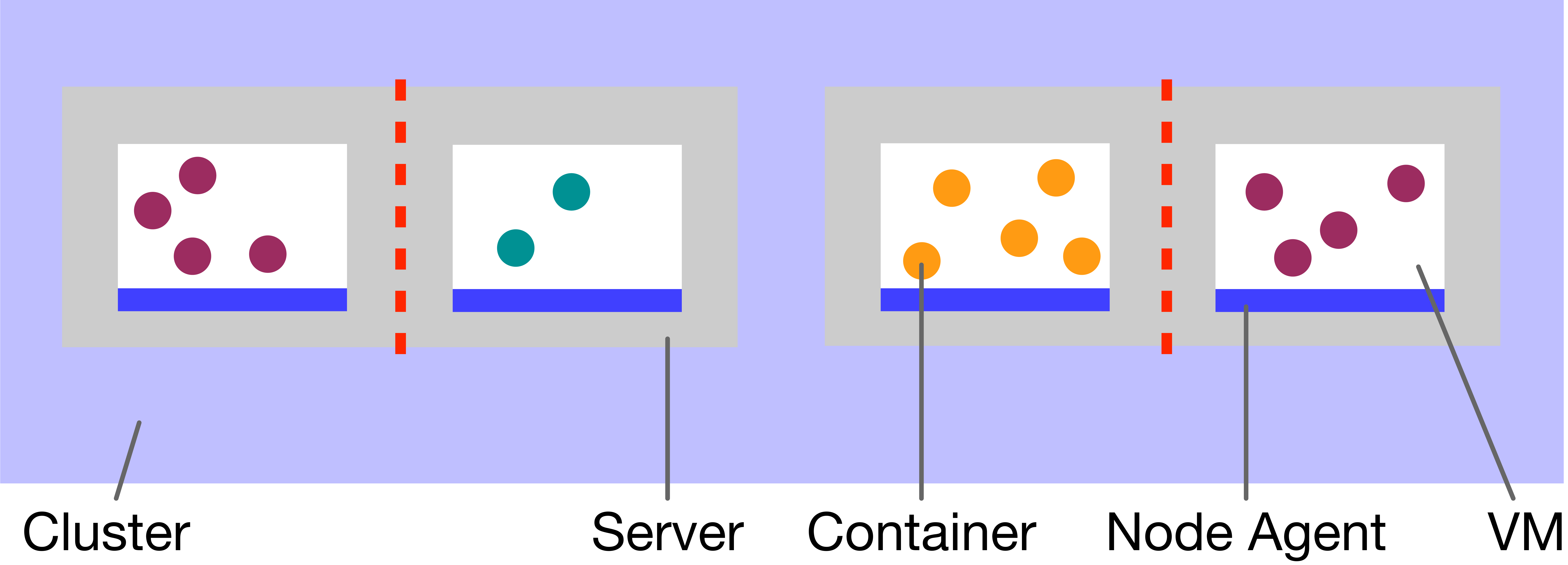}
        \label{fig:slicing-vms-node-level}
    } \\
    \subfloat[\textbf{Sub-node-level slicing of servers.} A subset of resources of a node that is a physical server is made available to a tenant.]{
        \includegraphics[width=0.42\linewidth]{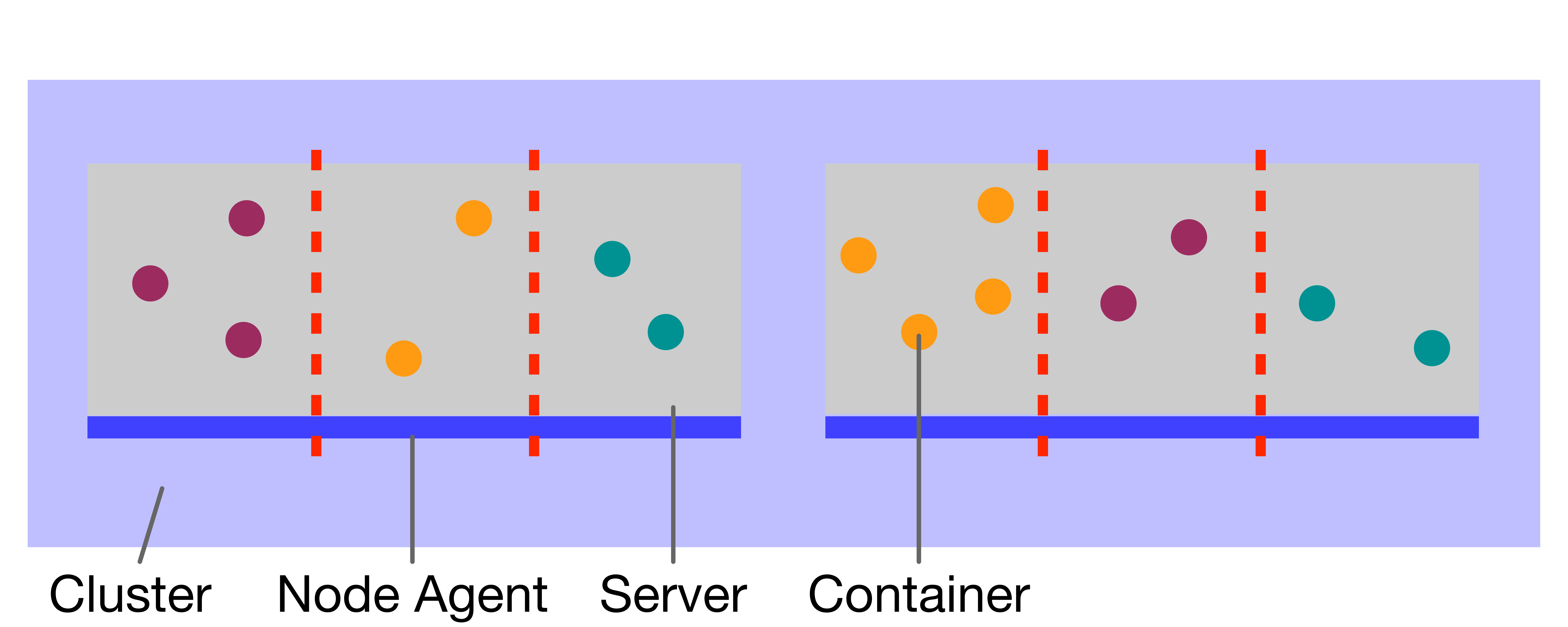}
        \label{fig:slicing-servers-sub-node-level}
    }
    \hspace{0.1\textwidth}
    \subfloat[\textbf{Sub-node-level slicing of VMs.} A subset of resources of a node that is a VM is made available to a tenant.]{
        \includegraphics[width=0.42\linewidth]{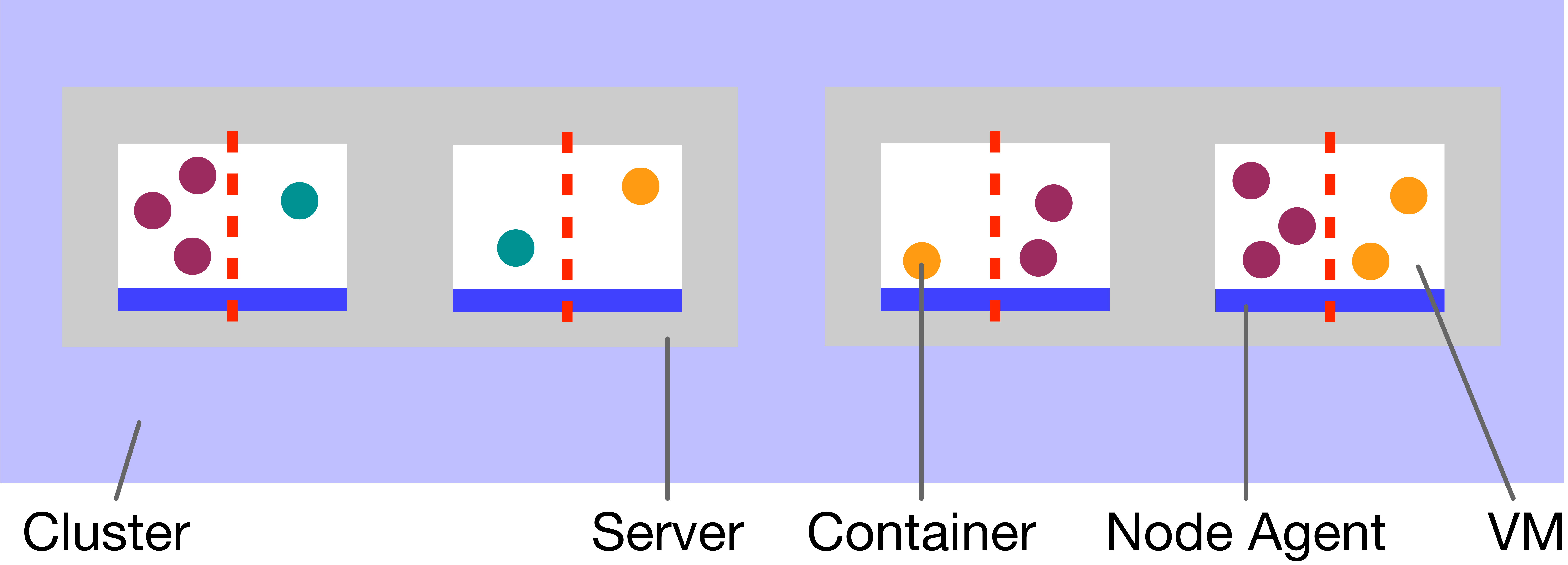}
        \label{fig:slicing-vms-sub-node-level}
    } 
    \vspace{14pt}
    \caption{\textbf{Node and slice granularities.} 
    Dashed vertical lines indicate how a cluster's resources are sliced so as to make those resources available to tenants.
    A node in a cluster can be a physical server (left illustrations) or a VM (right illustrations), presented as node granularities.
    Slicing can be performed so as to make an entire node available to a tenant (top illustrations) or so as to make a subset of a node's resources available to a tenant (bottom illustrations).
    Different node and slice granularities can coexist within a cluster (e.g., the scenarios shown in all four illustrations could appear simultaneously in a single cluster).
    Our EdgeNet multitenancy framework automates the process of varying the slice granularity, allowing a node to be reserved for a tenant, or returning a reserved node to the pool of nodes available to be subdivided.}
    \label{fig:slice-granularity}
\end{figure*}

We use the term \textit{slicing} to refer to a mechanism that enables multitenancy by dividing a larger pool of resources into smaller portions, each portion being for the exclusive use of one of the tenants.
For CaaS, the larger pool is a compute cluster that consists of nodes, which may be either physical servers or virtual machines.
But what size should a smaller portion be: a full node, or a subset of the resources of a node?
A subset can be acquired through the use of containers, sandboxed to a greater or lesser degree, as Sec.~\ref{sec-container-isolation} will describe.
Fig.~\ref{fig:slice-granularity} depicts the different possible node and slice granularities.
In our estimation, neither of the slicing granularities is ideal for all use cases, and a multitenancy framework should offer both, and automate the ability to switch between them. 

\textbf{Node-level Slicing} (Figs.~\ref{fig:slicing-servers-node-level} and \ref{fig:slicing-vms-node-level}).
Slicing at this granularity, which is offered by all of the frameworks that we have studied, provides a tenant with one or more entire nodes, so that isolation of a tenant workload is ensured at the level of the node in which it runs.
By this means, it offers greater freedom in choosing a container runtime to support a particular containerized workload.
And it can better ensure stable access to resources.
Reserving an entire physical server (Fig.~\ref{fig:slicing-servers-node-level}) can be valuable, in particular, for a tenant that needs to meet an unusual requirement, such as guaranteed access to GPU resources.
However, when entire nodes are reserved for tenants, some nodes might be under-utilized.

\textbf{Sub-node-level Slicing} (Figs.~\ref{fig:slicing-servers-sub-node-level} and \ref{fig:slicing-vms-sub-node-level}).
Sub-node-level slicing improves the ability of a cluster to maximize the efficiency of its resources.
This is enabled through containers where each container on a node takes a portion of its resources.
Isolation between multi-tenant workloads on the same host is provided at the level of containers, so it is weak.
Better isolation can be ensured through container runtimes that provide sandboxes to containers.
This approach restricts tenant autonomy in selecting a container runtime as there are just a few of them available.

As Table~\ref{tab:related-work-comparison} shows, all of the CaaS multitenancy frameworks that we have studied offer node-level slicing, and all but Kamaji offer sub-node-level slicing.
When it is available, sub-node-level slicing is the default.
Upon the request of a tenant, a cluster administrator can manually configure node-level slicing.

The EdgeNet framework is the only one for which the process of switching granularity is automated.
Sec.~\ref{sec-architecture-slice} describes how we implement this.
It might seem that the node-level slicing that we thereby enable suffers from all of the inefficiency of the multi-instance CaaS model that we have critiqued (see Sec.~\ref{sec-introduction}), but this is not so, as our architecture preserves the single-instance efficiency of a single control plane.

\subsection{Federation support}
\label{sec-federation-support}
% \label{sec-design-federation}
CaaS multitenancy frameworks have to date generally been aimed at the use case of a single cluster operator offering its resources to its own tenants.
However, the resources of several operators from different regions or countries will generally be required by a tenant that wishes to provide its edge cloud based services to large numbers of end-users.
Such a tenant might prefer to be the customer of just one operator and, through that operator, gain access to the others.
We anticipate that operators will see a commercial interest in federation, which will allow them to more broadly commercialize access to their clusters.
We also anticipate that operators will want to lower the barrier to entry for those who deploy services by allowing them to orchestrate their containers across multiple clusters with a single tool.

Many edge cloud services, such as cognitive services~\cite{ding2020cloudedge, gholami2021cloudedge}, are expected to involve workloads that are spread across the cloud and the edge cloud~\cite{abdul2019containeroffloading}, with workloads moving back and forth between the two, so there are voices in industry that argue~\cite{yi2015fogsurvey}, and we are convinced, that a unified, single interface for users is a necessity.
As a first step towards this goal, the EdgeNet multitenancy architecture presents an essential first brick in such a federation architecture: the ability to generate object names that are universally unique to cluster and tenant.
Such uniqueness avoids name collisions during the propagation of objects across clusters.
The details of our implementation are found in Sec.~\ref{sec-architecture-subnamespace-federation}.

Besides our EdgeNet framework, five of the frameworks that we study support scaling up the infrastructure that multiple tenants share, and four of them, the Virtual Kubelet based frameworks, do so through federation.
% (see Table~\ref{tab:related-work-comparison}).
Even for their main purpose of enabling deployment of workloads to multiple clusters, Virtual Kubelet based frameworks suffer from a significant drawback: Kubernetes' automatic scaling up and down of workloads to meet demand gets lost in remote clusters.
This is because the Kubernetes objects that get deployed through a virtual kubelet are pods rather than the \textit{Deployment} or \textit{StatefulSet} workload resources that manage pod life cycles on a user's behalf, and the Kubernetes Horizontal Pod Autoscaling mechanism\footnote{Kubernetes documentation: \textit{Horizontal Pod Autoscaling} \url{https://kubernetes.io/docs/tasks/run-application/horizontal-pod-autoscale/}} in each cluster works on these sorts of objects, not on individual pods.

Like Virtual Kubelet, EdgeNet enables the deployment of workloads from local clusters on remote clusters, but EdgeNet handles this through an intermediate cluster between local and remote clusters.
The intermediate cluster that does this for EdgeNet is called the \emph{Federation Manager}.
When a tenant, using its local cluster, makes a deployment in federation scope, the Federation Manager creates the deployment on the remote cluster on behalf of the tenant, as we describe briefly in Sec.~\ref{sec-architecture-subnamespace-federation}.

Some of Liqo's extensions to Virtual Kubelet start to tackle some of the concerns that would arise in a multi-tenant federation, such as collisions between the names of namespaces generated in local clusters and in remote clusters.
Liqo's solution is a naming scheme that ensures that the name of a namespace used by a workload will be unique in the remote cluster in which it is deployed.\footnote{Liqo documentation: \textit{Namespace Offloading} \url{https://docs.liqo.io/en/v0.7.0/usage/namespace-offloading.html}}
However, the same workload risks running in namespaces with different names in different clusters, which can itself lead to problems.
EdgeNet by contrast generates globally unique names that avoid collisions, and a workload runs in namespaces that carry the same name on all clusters to which it is deployed. 

The other framework that provides for cloud-edge communication and significant scaling is Arktos, but we have been unable to determine whether federation is involved.
Its stated aim is to achieve a single regional control plane to manage 300,000 nodes that multiple tenants will share.\footnote{Arktos documentation: \emph{Large Scalability} \url{https://github.com/CentaurusInfra/arktos\#large-scalability}}

\section{Design Decisions}
\label{sec-design}
Our vision for EdgeNet's multitenancy framework is to promote a future in which the CaaS service model can thrive, particularly at the network edge.
We have made nine design decisions, listed below, to support this vision.
The first six were discussed in relation to related work in the previous section, and the latter three are discussed in this section.
The implementation details are provided in the Architecture section that follows (Sec.~\ref{sec-architecture}).

\begin{itemize}
    \item \textbf{Multitenancy approach.} EdgeNet obtains the lower overhead offered by a \textit{single-instance native} approach to multitenancy, compromising on the isolation that would be offered by a \textit{multi-instance} one (Sec.~\ref{sec-multitenancy-approach}).

    \item \textbf{Customization approach.} We mitigate customization limitations that stem from the single-instance approach through the use of hierarchical namespaces (Sec.~\ref{sec-tenant-environment-customization}).

    \item \textbf{Consumer and vendor tenancy.} We design EdgeNet to support both the \textit{consumer} and \textit{vendor} forms of tenancy (Sec.~\ref{sec-consumer-and-vendor-modes}).

    \item \textbf{Tenant resource quota.} EdgeNet incorporates a control mechanism to manage the allocation of resource quotas in a hierarchical tenancy structure, allowing tenants to grant quotas to their subtenants and recoup those quotas from them (Sec.~\ref{sec-tenant-resource-quota}).

    \item \textbf{Variable slice granularity.}
    Considering that there is no ideal granularity at which to slice a compute cluster in order to deliver resources to tenants, we allow an EdgeNet cluster to be sliced %either 
   into individual compute nodes or at a sub-node-level granularity (Sec.~\ref{sec-variable-slice-granularity}).%or, more finely grained, into individual virtual machines (Sec.~\ref{sec-design-workload-isolation}).

    \item \textbf{Federation support.} Our framework allows each EdgeNet cluster to receive the workloads of tenants from other EdgeNet clusters with which it is federated, while avoiding name collisions by generating object names that are unique to cluster and tenant (Sec.~\ref{sec-federation-support}).

    \item \textbf{Kubernetes custom resources.} For ease of integration into existing systems and ease of adoption by users, we implement EdgeNet using the Kubernetes \textit{custom resources} feature, rather than creating a wrapper around Kubernetes or forking the Kubernetes code (Sec.~\ref{sec-design-kubernetes}).

     \item \textbf{Lightweight hardware virtualization.} We compensate for the loosened isolation of workloads in the single-instance native approach through the use of lightweight hardware virtualization that is optimized for running containers (Sec.~\ref{sec-container-isolation}).  

     \item \textbf{External authentication.} In a federated multitenancy environment, users will need to authenticate with remote clusters, and for that reason EdgeNet adopts an authentication method that is external to any individual cluster (Sec.~\ref{sec-design-authentication}).
 
\end{itemize}

\subsection{Kubernetes custom resources}
\label{sec-design-kubernetes}
Kubernetes' custom resource feature\footnote{Kubernetes documentation: \textit{Custom Resources} \url{https://kubernetes.io/docs/concepts/extend-kubernetes/api-extension/custom-resources/}} allows new entities to be added that, by the fact of their presence, extend the standard Kubernetes API, thereby maintaining backward compatibility with tools and interfaces that are familiar to users.
By building our EdgeNet framework in this way, instead of as a wrapper around Kubernetes or as a separate system that interacts with Kubernetes, we increase the chances that the framework will be compatible with a variety of Kubernetes distributions.
For example, we have successfully tested and run EdgeNet framework as an extension of k3s,\footnote{K3s \url{https://k3s.io/}} a lightweight certified Kubernetes distribution for IoT and edge computing.

We have containerized the EdgeNet extensions and we provide them in the form of public Docker images and configuration files.
The core Kubernetes code remains untouched, and there is no need to recompile any existing code that runs a cluster.
Any cluster administrator can deploy the extensions to their cluster with a single \emph{kubectl apply} command without the need to bring down the cluster or interrupt its work in any way.

Aside from the choice of Kubernetes and of Kubernetes custom resources, all of our other design decisions should, in principle, apply to enabling multitenancy in any other container orchestration tool.

\subsection{Lightweight hardware virtualization}
\label{sec-container-isolation}

The choice of virtualization technology, in the context of edge computing, between hypervisors providing the best isolation and containers being lightweight~\cite{yi2015fogcomputing}, is a longstanding discussion. 
We prioritize virtualized environments because of their lower overhead; in so doing, we favor enhanced performance over delivering the best isolation~\cite{julien2019edgesurvey}.
A native framework with operating-system-level virtualization satisfies these requirements, but it presents security concerns having to do with containers sharing the same kernel. %incurs low overhead
We want to offer each tenant the security of its own guest kernel, which hardware virtualization provides, but without going so far as to adopt a multi-instance approach that would negate the performance advantages of containers over VMs.
Fortunately, this is possible through the use of lightweight virtual machines, which offer the isolation benefits of hardware virtualization while offering near-container-level performance.
Our multitenancy framework therefore adopts a single-instance native approach with lightweight hardware virtualization.

We follow earlier work~\cite{flauzac2020review, randazzo2019kata} that has recommended the Kata runtime\footnote{Kata containers \url{https://katacontainers.io/}} for providing isolation between containers in a multitenant environment~\cite{de2020understanding, kumar2020performance, aggarwal2020performance, viktorsson2020tradeoff}.
Kata spawns a lightweight VM that is optimized to run containers, delivering near-container-level performance~\cite[Fig.~5]{de2020understanding} and better isolation than OS-level virtualization.

\begin{figure*}
    \centering
    \subfloat[One workload per virtual machine, the best isolation among the three, introducing high overhead.]{
        \includegraphics[width=0.3\linewidth]{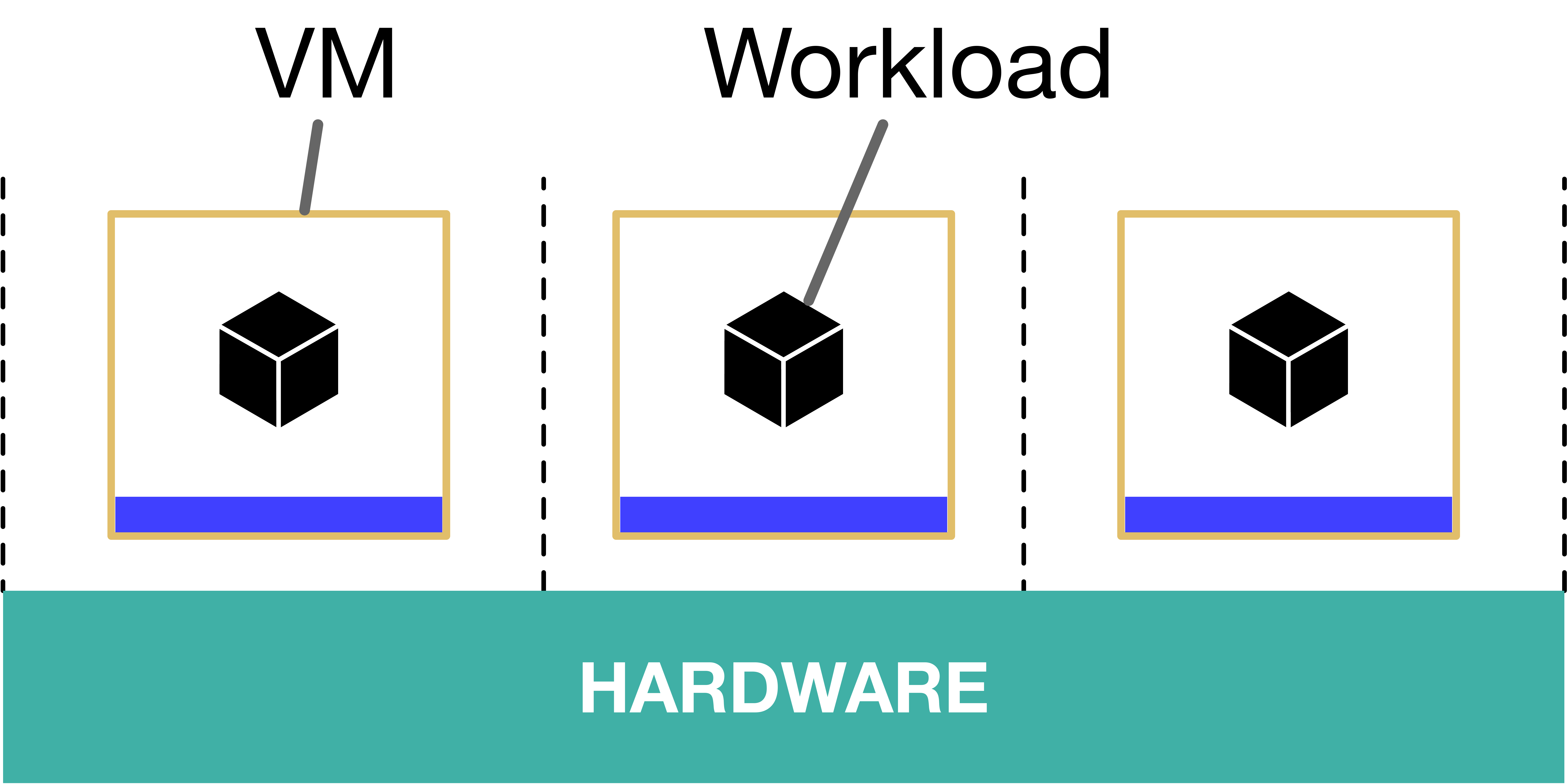}
        \label{fig:workload-isolation-vm}
    }
    \hfill
    \subfloat[One workload per container, the weakest isolation among the three, providing improved performance.]{
        \includegraphics[width=0.3\linewidth]{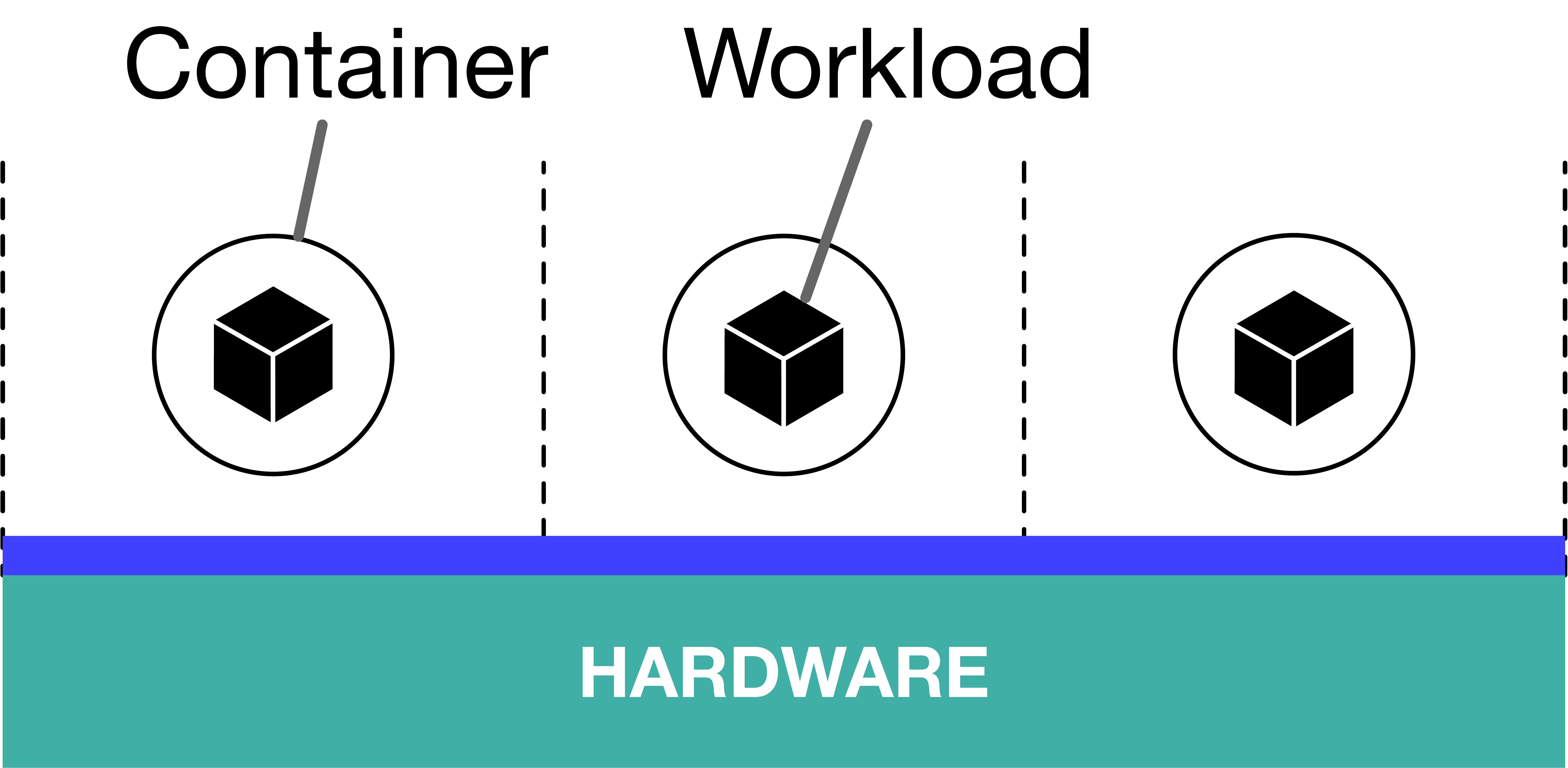}
        \label{fig:workload-isolation-containers}
    }
    \hfill
    \subfloat[One workload per container sandboxed by Kata, providing both intermediate isolation and intermediate overhead.]{
        \includegraphics[width=0.3\linewidth]{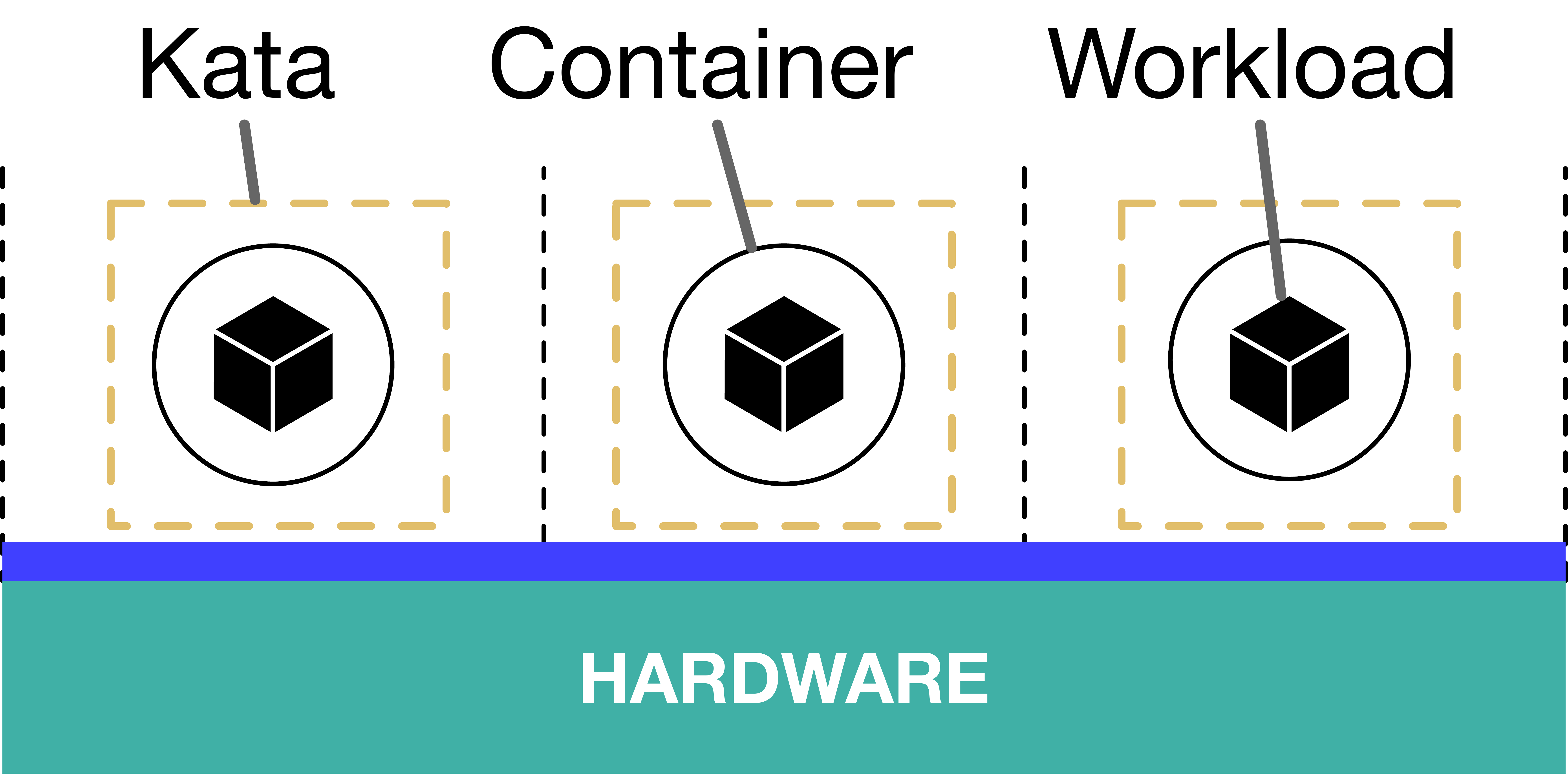}
        \label{fig:workload-isolation-kata}
    }
    \caption{Methods for isolating workloads.
    The dashed vertical lines distinguish one tenant from another.
    Each thick blue horizontal line designates a node.}
    \label{fig:choice-of-workload-isolation}
\end{figure*}

Fig.~\ref{fig:choice-of-workload-isolation} depicts three methods for workload isolation: virtual machines, Docker containers, and Kata containers.
We consider a single workload per method that can improve isolation and performance at the cost of overhead.
One workload per virtual machine provides the best isolation among the three while introducing high overhead.
The containerization technique can lower such overhead, having one workload per container, although it diminishes the isolation.
The Kata method falls between VMs and containers in terms of isolation and overhead, as a containerized single workload runs in a lightweight virtual machine.

Tenants who require better isolation and performance at the same time, can obtain these using the \textit{slice} software entity in our framework.
As described in Sec.~\ref{sec-architecture-slice}, this entity provides a tenant with the option of selecting container runtimes on an isolated subcluster so that the tenant can select one that meets its application requirements.

\subsection{External Authentication}
\label{sec-design-authentication}

A tenant's users must authenticate themselves in order to access the resources that they are authorized to access.
For multitenant CaaS to run at scale, it is not feasible to require users to have individual accounts at every different cluster location where they will deploy their workloads~\cite{bohn2020nistfederation}.
Instead, authentication should be managed by an integrated identity management system.
For example, an identity federation that consists of multiple identity providers, using OpenID Connect (OIDC)\footnote{OpenID Connect \url{https://openid.net/connect/}} running on top of OAuth 2.0\footnote{OAuth 2.0 \url{https://oauth.net/2/}} as the authentication method, can support large-scale federations.
With this in mind, EdgeNet uses this type of authentication (See Sec.~\ref{sec-architecture-authenication}).

\section{Architecture}
\label{sec-architecture}
\begin{figure*}[t]
  \centering
  \includegraphics[width=\linewidth]{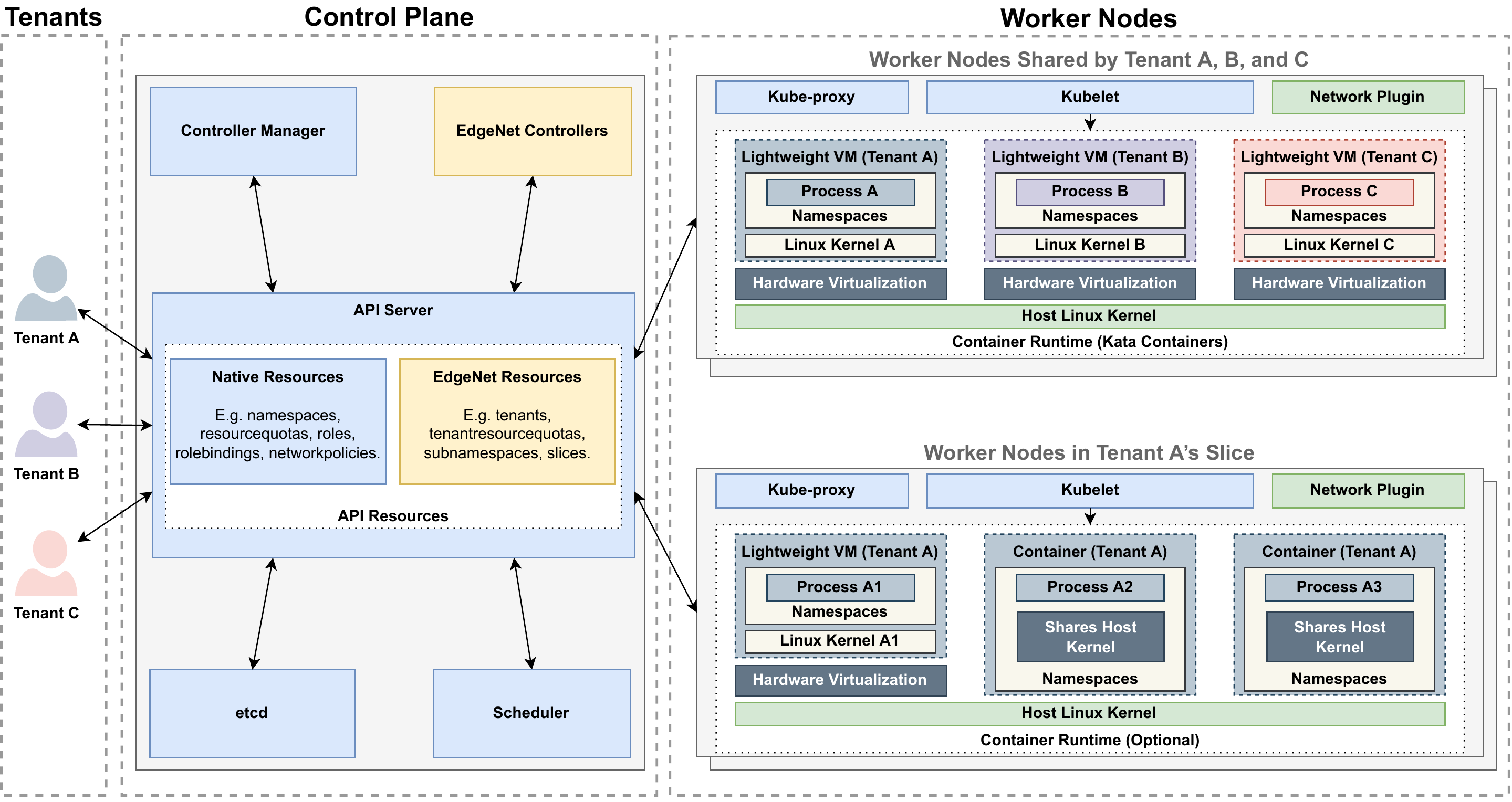}
  \caption{Architectural overview of EdgeNet multitenant CaaS framework implemented on Kubernetes.
  Much of the architecture is built upon native Kubernetes (blue) and third-party software components (green).
  Our innovation can be seen in the control plane, which consists of resources and controllers (yellow).
  This allows multiple tenants (left) to make use of the same control plane (center) via the API server to create workloads (right).
  What is more, our contributions enable the creation of two types of worker nodes; shared (top right) and reserved to a node-level slice (bottom right).
  Upon inclusion in a slice, a worker node's type switches to reserved, reverting to shared after slice termination.
  Multitenant workloads (top right) share the compute resources of shared worker nodes, yet, each is isolated from another through hardware virtualization, lightweight virtual machines that are optimized for running containers, called Kata Containers.
  Every pod has its lightweight VM-based sandbox for isolation, and container(s) defined in a pod specification runs in the virtual machine tied to that pod. 
  This single instance shared approach eliminates an overhead introduced due to employing conventional virtual machines for the isolation of single-tenant clusters from each other, addressing the overhead not only related to worker nodes but also control plane.
  With that being said, worker nodes in a slice that is dynamically created by a tenant (bottom right) are isolated from multitenant workloads, hence providing the tenant with container runtime selection.
  In this way, the tenant can make the most of the advantages of containerization, such as lower overhead, and shorter creation and startup time.
  At the bottom right of the figure, several worker nodes are subclustered in Tenant A's slice, each hosting the workloads, for which the tenant employs Kata Containers as well as another container runtime like runC. 
  }
  \label{fig:architecture}
\end{figure*}

Our EdgeNet architecture has been conceived around the design decisions articulated in Sec.~\ref{sec-design}, with the aim of introducing as low overhead as possible while making Kubernetes ready for the edge. % and reduce costs.
As a reminder, our main design decision has been to take a single-instance native approach, meaning that tenants share a cluster's control plane components and compute nodes, rather than having each tenant acquire its own control plane components and compute nodes.
To compensate for the diminished isolation that comes with sharing the same cluster, EdgeNet uses lightweight VMs to isolate workloads while retaining low overhead. 

The architecture of our EdgeNet multitenant CaaS framework is illustrated in Fig.~\ref{fig:architecture}. 
It is designed as a set of custom resources and custom controllers that extend Kubernetes from within.
The framework consists of six principal new entities:\footnote{EdgeNet multitenancy software entities: Principal custom controllers \url{https://github.com/EdgeNet-project/edgenet/tree/v1.0.0-alpha.5/pkg/controller/core/v1alpha1}, Assistant custom controllers \url{https://github.com/EdgeNet-project/edgenet/tree/v1.0.0-alpha.5/pkg/controller/registration/v1alpha1}, Admission control webhook \url{https://github.com/EdgeNet-project/edgenet/tree/v1.0.0-alpha.5/pkg/admissioncontrol}}
\begin{itemize}
    \item \textit{Tenant} is the fundamental entity that isolates a tenant from other tenants (Sec.~\ref{sec-architecture-tenant}).
    \item \textit{Subsidiary Namespace} is an isolated environment created by a tenant (Sec.~\ref{sec-architecture-subnamespace}).
    \item \textit{Tenant Resource Quota} controls a tenant's use of resources (Sec.~\ref{sec-architecture-trq}).
    \item Two entities, \textit{Slice} and \textit{Slice Claim}, allow dynamically reserving sub-clusters isolated from multitenant workloads, entitled node-level-slicing (Sec.~\ref{sec-architecture-slice}).
    \item \textit{Admission Control Webhook} enforces custom policies\footnote{Kubernetes documentation: \textit{Dynamic Admission Control} \url{https://kubernetes.io/docs/reference/access-authn-authz/extensible-admission-controllers/\#what-are-admission-webhooks}} such as employing Kata Containers for multitenant workloads (Sec.~\ref{sec-architecture-admission-webhook}).
\end{itemize}
These are assisted by new entities that facilitate cluster and tenant management: \textit{Role Request} (Sec.~\ref{sec-architecture-rolerequest}), and \textit{Tenant Request} and \textit{Cluster Role Request} (Sec.~\ref{sec-architecture-others}).
Our architecture also covers user authentication via existing mechanisms (Sec.~\ref{sec-architecture-authenication}).
Aside from these, it provides cluster operators with configuration files in YAML format that can be carefully customized, which define runtime class\footnote{Kubernetes documentation: \textit{Runtime Class} \url{https://kubernetes.io/docs/concepts/containers/runtime-class/}} and predefined role resources.

\subsection{Tenant}
\label{sec-architecture-tenant}
\begin{figure}
    \centering
    \includegraphics[width=0.7\columnwidth]{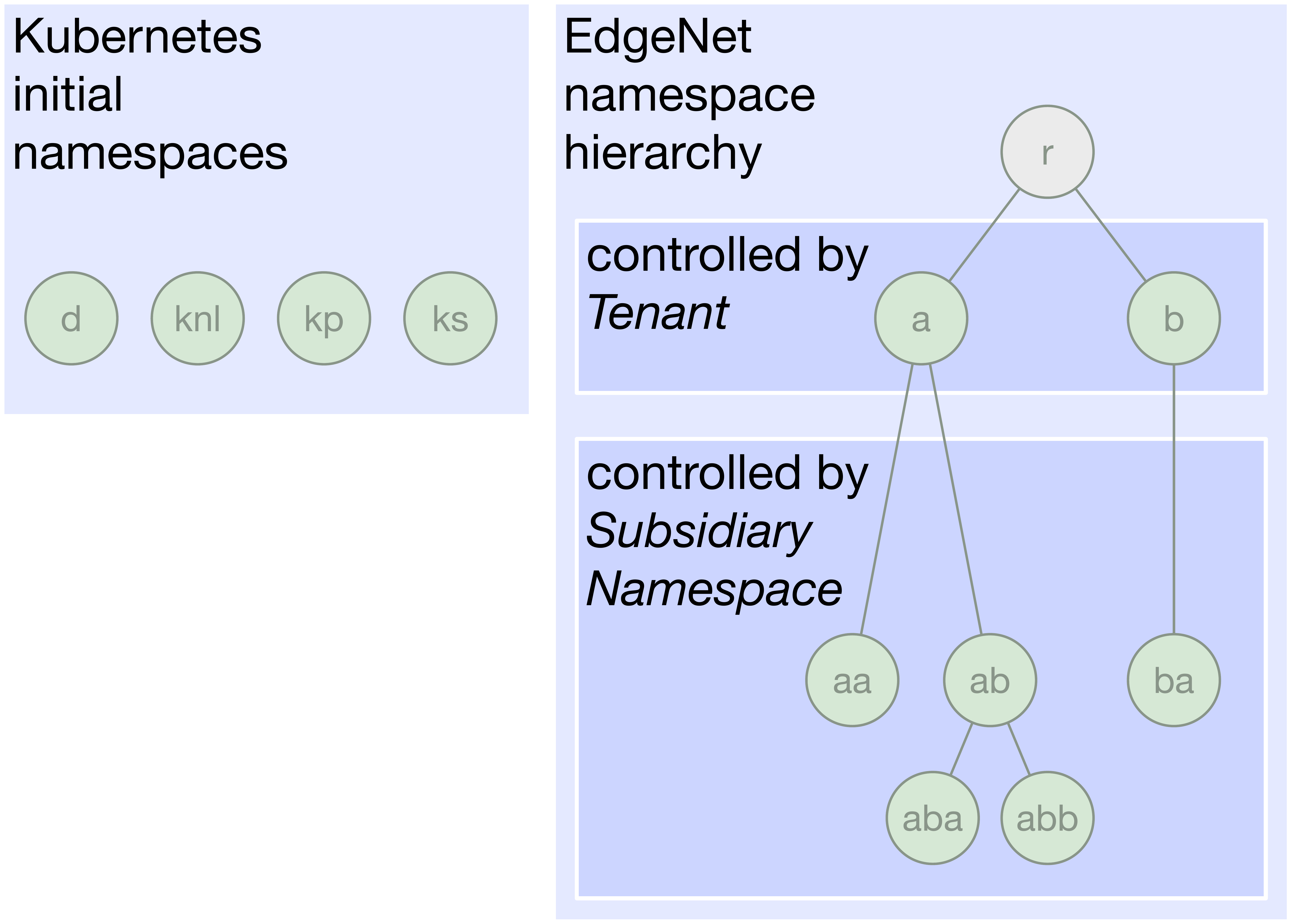}
    \caption{\textbf{Namespace Hierarchy in EdgeNet.}
    EdgeNet's multitenancy framework provides two principal controllers for managing its namespace hierarchy.
    The \textit{Tenant} controller creates, updates, and deletes the tenant core namespaces at the top level of the hierarchy, while the \textit{Subsidiary Namespace} controller handles all namespaces further down in the hierarchy.
    \newline \newline In this example, \textit{a} and \textit{b} are tenant core namespaces, directly under the root of the hierarchy, \textit{r}, which is not itself a namespace; the subsidiary namespaces are \textit{aa}, \textit{ab}, \textit{aba}, \textit{abb}, and \textit{ba}.
    \newline \newline Kubernetes' initial namespaces, \texttt{default} (\textit{d}), \texttt{kube-node-lease} (\textit{knl}), \texttt{kube-public} (\textit{kp}), and \texttt{kube-system} (\textit{ks}) are not included in the hierarchy and are not managed by these controllers.}
    \label{fig:namespace_hierarchy_edgenet}
\end{figure}

In the context of the namespace structure maintained by the EdgeNet framework, the \textit{Tenant} entity is a controller that acts at the top level of the hierarchy: creating, updating, and deleting the core namespaces of cluster-scoped tenants, which are the ones that are admitted into the cluster by the cluster's administrator.
Here, we describe the \textit{Tenant} entity, while Sec.~\ref{sec-architecture-subnamespace} describes the \textit{Subsidiary Namespace} entity, which acts lower down in the hierarchy, on the subtenants that are admitted either by top-level tenants or, recursively, by subtenants.
The roles of these two controllers are shown in Fig.~\ref{fig:namespace_hierarchy_edgenet}.

The \textit{Tenant} entity handles the creation of a tenant environment in a cluster.
It starts by generating a namespace for this tenant, using a tenant name supplied by the tenant and checked for uniqueness among the cluster's namespaces.
Because the tenant will be able to create its own hierarchy of namespaces rooted at this namespace, we distinguish this one, which will be at the root of any sub-tree that the tenant creates, by calling it the tenant's \textit{core namespace}.

The controller also applies four labels to the namespace:
\begin{itemize}
    \item \textit{kind=$<$namespace-type$>$}, which is \textit{core} in this case;
    \item \textit{tenant=$<$tenant-name$>$}, the name supplied by the tenant;
    \item \textit{tenant-uid=$<$tenant-uid$>$}, a locally-generated unique identifier for the tenant; 
    \item \textit{cluster-uid=$<$cluster-uid$>$}, the UID of the \textit{kube-system} namespace.
\end{itemize}
UIDs are defined in Kubernetes as being 128-bit-long universally unique identifiers~\cite{leach2005universally},\footnote{Kubernetes documentation: \textit{Object Names and IDs; UIDs} \url{https://kubernetes.io/docs/concepts/overview/working-with-objects/names/\#uids}} and the Kubernetes community suggests using the UID of the \textit{kube-system} namespace as a cluster identifier.\footnote{Cluster ID API discussion in the Kubernetes Architecture SIG mailing list \url{https://groups.google.com/g/kubernetes-sig-architecture/c/mVGobfD4TpY/m/uEjVVsinAAAJ}}
The labels allow the tenant namespaces to be consumed by policies and other entities locally.
This labeling model is also required for the inter-cluster object propagation mechanism.

Each tenant has an owner who has control over the tenant and its resources, including any subnamespaces that the tenant might create.
Having created the core namespace, the \textit{Tenant} entity uses the Kubernetes role-based access control (RBAC) mechanism to grant this control, while at the same time limiting the tenant owner's control to the scope of its core namespace, so that it may not interfere with other tenants' namespaces.
The \textit{Subsidiary Namespace} entity will be responsible for extending the scope of the owner's control to the subnamespaces.
With their control over the core namespace, the owner can manage the tenant by, among other things: admitting users; granting roles, which are sets of permissions, for those users; and deploying workloads.

Kubernetes' network policies allow confining pod communication into a namespace or set of namespaces by using labels.
In our multitenancy framework, the policies consume the UID labels, as specified earlier, attached to tenant namespaces.
Since tenants have complete authorization on their network policies, an authorized user can, wittingly or not, misconfigure network policies in a namespace, thus resulting in security threats.
To overcome this vulnerability, we let a tenant enable or disable cluster-level network policy in the tenant specification, which confines the tenant's namespaces thanks to VMware's Antrea.\footnote{Antrea \url{https://antrea.io/}}

\subsection{Subsidiary Namespaces}
\label{sec-architecture-subnamespace}
Authorizations are issued hierarchy-wise, establishing a chain of accountability.
In other words, the permissions of a tenant owner to use the system are granted in the tenant's core namespace, applying to all its hierarchical namespaces.
Each individual user in a tenant, in turn, is authorized by the tenant owner.
According to permissions granted, the owner can create different roles in different subnamespaces as needed.
For example, an owner can grant some users administrative rights to approve other users in core and subnamespaces.
As their permissions are limited to their hierarchy tree, tenants cannot interfere with other tenants' environments. 
A tenant, at the same time, can use the system as if it has the authorization to create namespaces directly, thus having a relatively customizable environment.

% The term \textit{subsidiary namespace}, 
The \textit{subsidiary namespaces} custom resource, also known as \textit{subnamespaces}, %is used to describe an entity that empowers tenants to 
is a software entity through which tenants can create Kubernetes namespaces without having the authorization to do so directly.
Subnamespaces are indispensable for realizing the key features of our framework that are described in this paper's introduction, as we see in our discussion of Modes, Inheritance, Naming Convention, and Federation, below.

\subsubsection{Consumer and Vendor Tenancy}
\label{sec-architecture-subnamespace-modes}    
\begin{figure*}[hbtp]
  \centering
  \includegraphics[width=0.5\linewidth]{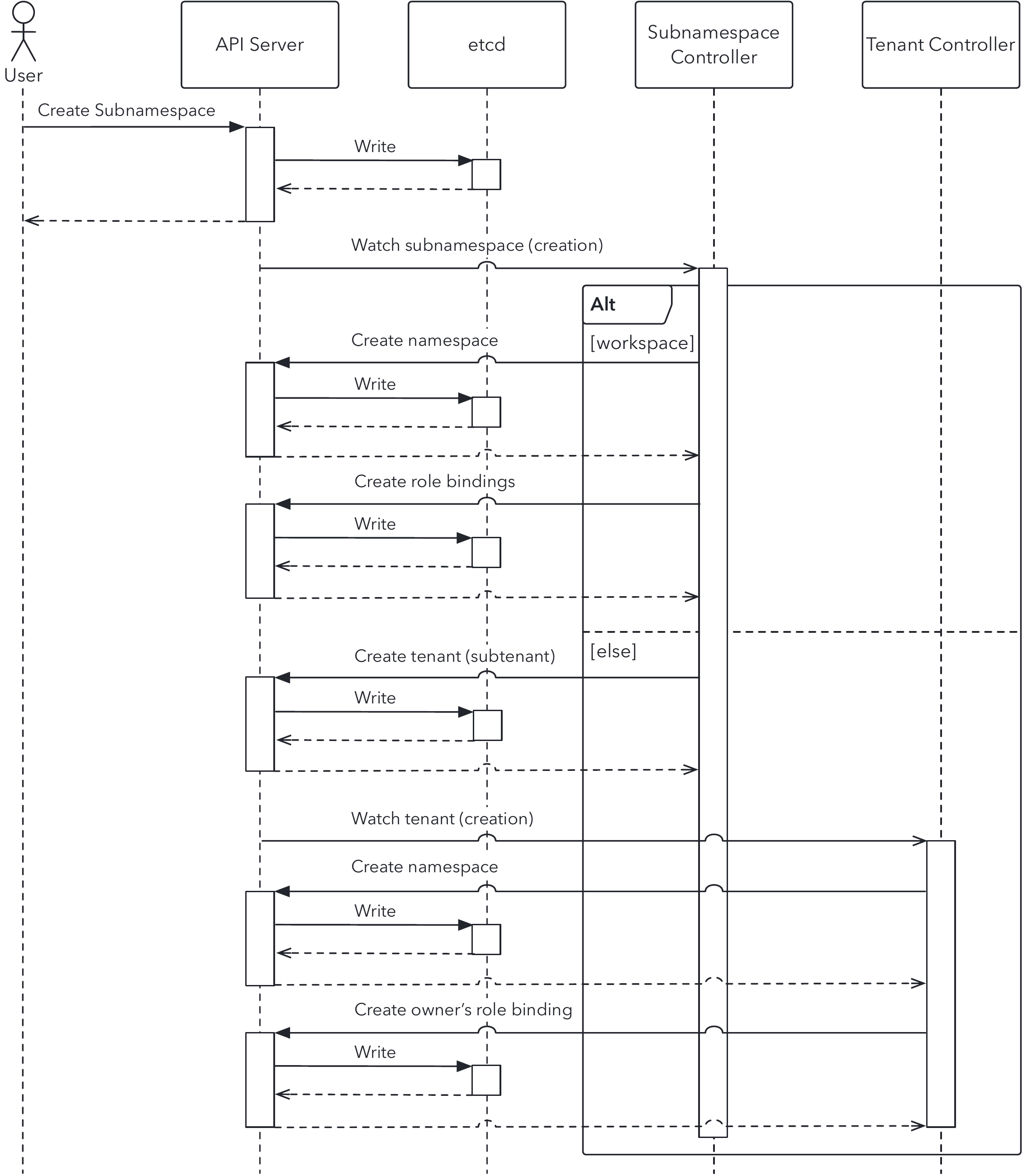}
  \caption{Sequence diagram of the \textit{Subnamespace} entity performing namespace and permission creation.
  It highlights how workspace and subtenant modes differ in doing so.}
  \label{fig:subnamespace-diagram}
\end{figure*}

The subnamespace entity relies on the parent-child relationship between the namespaces, starting from the core namespace of a tenant.
Each subsidiary namespace can be both a parent and a child at the same time.
The entity exists in one of two modes, \textit{workspace} or \textit{subtenant}, corresponding to the two forms of tenancy: consumer and vendor.
The sequence diagram in Fig.~\ref{fig:subnamespace-diagram} sketches out how the workspace and subtenant modes differ in creating a child namespace.

The hierarchical namespaces approach allows an organization to isolate workspaces for different products by generating a subnamespace per product. 
Each child namespace assigned to a product can be used to isolate its teams, for example, backend and frontend, in the same way.
Another real-world example consists in using subnamespaces to create isolated environments for multiple groups of students working on a laboratory experiment.

\begin{figure}
    \centering
    \includegraphics[width=0.7\columnwidth]{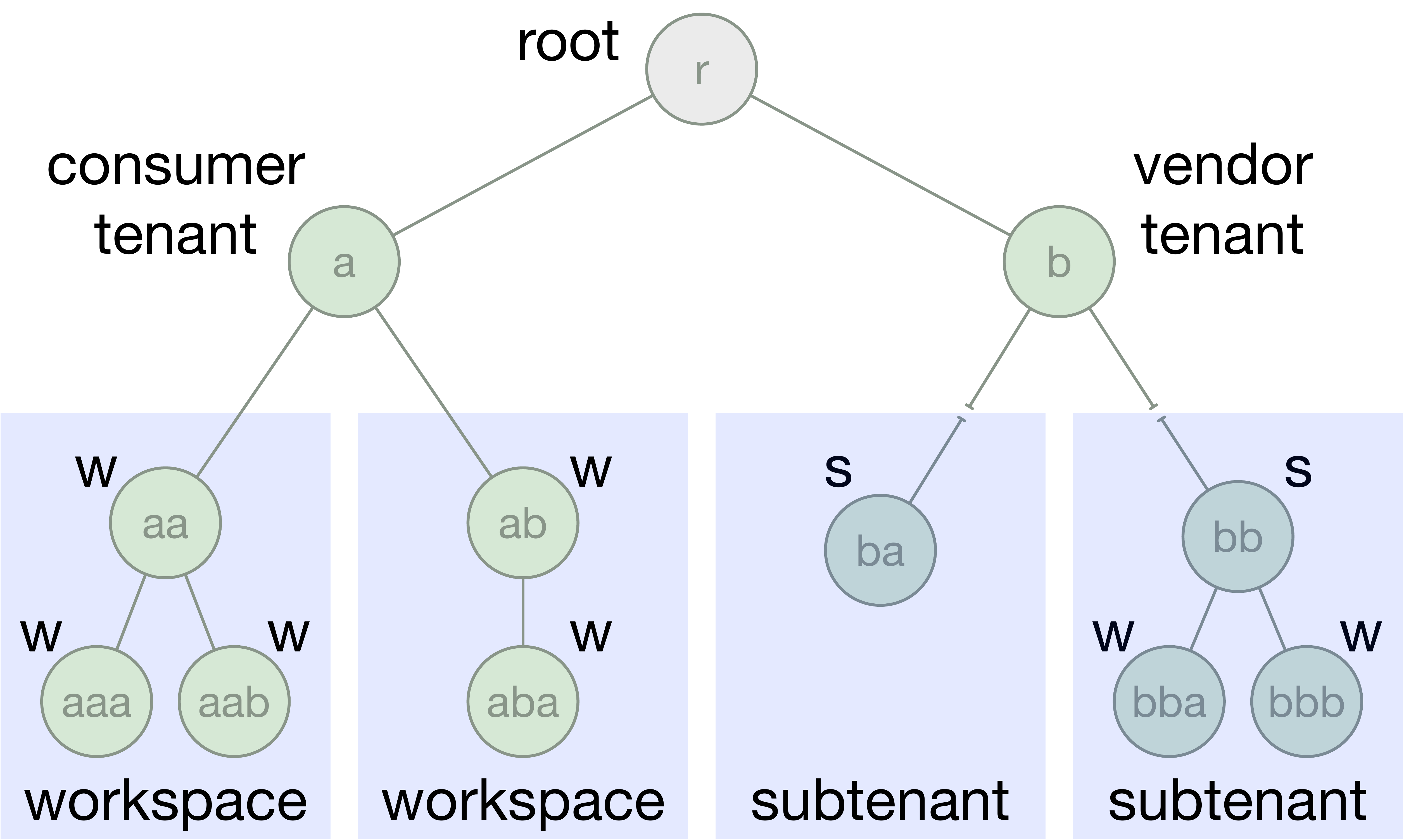}
    \caption{\textbf{Consumer and Vendor Tenancy in EdgeNet, showing workspace (w) and subtenant (s) modes.} 
    EdgeNet uses its hierarchical namespace structure to build consumer and vendor tenancy.
    In this example, the namespace \textit{a} belongs to a consumer tenant and the namespace \textit{b} belongs to a vendor tenant that is reselling containers\--as\--a\--service to its own customers.
    \newline \newline The \textit{Subnamespace} controller creates workspaces rooted at \textit{aa} and \textit{ab} for the \textbf{consumer tenant} by placing those namespaces into \textit{workspace} mode.
    The consumer tenant has visibility into those workspaces.
    \newline \newline The controller creates subtenants rooted at \textit{ba} and \textit{bb} for the \textbf{vendor tenant} by placing those namespaces into \textit{subtenant} mode.
    The vendor tenant does not have visibility into its subtenants.
    Note that the subtenant that owns the sub-tree rooted at \textit{bb} does have visibility into its own workspaces at \textit{bba} and \textit{bbb}.
    }
    \label{fig:consumer_and_vendor_EdgeNet}
\end{figure}

Fig.~\ref{fig:consumer_and_vendor_EdgeNet} demonstrates a tree of namespaces where a parent is blind to information about a child's namespace and its children.
This shielding assists a tenant in subleasing the desired amount of resources to its customers, thus becoming a vendor.
Accordingly, the customer of a vendor becomes a \textit{subtenant}.
A vendor can remove any of its subtenants when the customer-vendor relationship comes to an end.

A key characteristic of subnamespaces is enabling the choice of either mode, workspace or subtenant, at any depth in the hierarchy.
By extension, subnamespaces allow a subtenant to be created in a child namespace with the workspace mode and another to be created with the subtenant mode, as shown in Fig.~\ref{fig:consumer_and_vendor_EdgeNet}.
Not only can these two modes co-exist in the same subtree, but they also reinforce each other's benefits.
Last but not least, a subsidiary namespace can also be formed to be propagated across federated clusters.
If so, it generates object names that are unique to the originating cluster and tenant to prevent name collisions during object propagation across the federation.
Sec.~\ref{sec-architecture-subnamespace-federation} describes how our federation architecture functions.

\subsubsection{Inheritance}
In the subnamespace specification, an authorized user can declare which objects are passed by inheritance from parent to child.
The Kubernetes resource kinds that can be inherited are currently as follows:
\begin{itemize}
    \item Role-based access control (RBAC): Roles and Role Bindings; both together adjust permissions of users.
    \item Network policies; make a namespace restricted to defined ingress/egress rules.
    \item Limit ranges; set a resource quota per pod.
    \item Secrets; keep sensitive information such as credentials to be consumed by pods.
    \item Config Maps; configuration to be used by pods.
    \item Service Accounts; an entity that allows applications and services to authenticate with the Kubernetes API.
\end{itemize}
If RBAC objects are not inherited, the specification must include the owner of the subnamespace for management purposes.
Further, it is possible to declare continuous inheritance.
In this case, the controller constantly syncs objects from a parent to its child.

Note that a resource quota is not an entity subject to inheritance, so as to avoid overconsumption by a tenant, which could get around quotas by generating subnamespaces at will.
The logic ensures that the aggregated child resource quotas cannot exceed their parent's initial resource quota, including the core namespace. 
Each subnamespace creation taxes its parent's resource quota so that the aggregation of resource quotas in the parent and child namespaces remain the same.
In other words, a tenant's resource quota is a cake to be shared out, and each subnamespace gets a piece of cake from its parent's cut.

\subsubsection{Naming Convention}
\label{sec-architecture-naming-convention}

The naming convention has been conceived so as to enable federation deployments.
As mentioned in Sec.~\ref{sec-architecture-tenant}, a core namespace shares the same name with its tenant.
Independent of its depth, a subnamespace follows the pattern of \textit{$<$subnamespace-name$>$-$<$hash$>$}.
We feed the hash function with the parent namespace and subnamespace name.
This naming convention reduces the chance of name collisions while creating subnamespaces.
If a collision nonetheless occurs, the subnamespace object enters a failure state, indicating a collision status.
This is vital to the interoperability of multiple clusters.
The reason is that tenants or namespaces holding the same names in different clusters probably occur in many clusters.
Consequently, conflicts will inevitably arise while propagating objects, unless there is an adjustment mechanism such as the one described here.

\subsubsection{Federation}
\label{sec-architecture-subnamespace-federation}

In our federation vision, each cluster, even before it is federated, is a multitenant cluster, making its worker nodes available to multiple tenants, and federation further opens the cluster to the workloads of tenants from other clusters.
(As we have discussed in Sec.~\ref{sec-related-work}, this differs from the approach of the Liqo framework, based on Virtual Kubelet, in which clusters only achieve multitenancy by federating.)
We have developed a proof-of-concept federation architecture with a prototype implementation, which works jointly with our multitenancy framework.
The source code of the prototype is publicly accessible via our repository.

We see each tenant gaining access to a federated set of clusters via what we might term a home cluster or local cluster.
For example, a company that has developed an application that serves vehicles in several countries might need to deploy its workloads to the edge clusters of mobile operators in each of those countries, and it can do so via a cluster in its home country that is federated with these other clusters.
To obtain access to a local cluster, it might contract with a cloud provider that has a commercial presence in its home country, leaving the cloud provider to manage the commercial relationships with the other providers in the federation.

Information regarding the identity of the company and its contract with its local provider remains local, while only the workload-related objects necessary for the deployment of the application get propagated to remote clusters.
Propagating as few objects as possible has three significant benefits: (1) it avoids replication of tenant information across clusters, thus reducing bandwidth consumption and unnecessary traffic; (2) it enhances data privacy and sovereignty and mitigates security risks; and (3) it significantly reduces overhead that could stem from running a control plane or worker nodes per tenant at the scale of a federation. 

In EdgeNet, the deployment scope of any subnamespace can be set to either \textit{federated} or \textit{local}.
If federated, the subnamespace controller adds the UID of the kube-system namespace as a prefix to the namespace name, and this cluster UID is also fed into the hash function described just above (Sec.~\ref{sec-architecture-naming-convention}).
This ensures the uniqueness of each name across all of the federated clusters.

In our prototype federation, a tenant deploys its workloads to remote clusters by creating a \textit{Selective Deployment}~\cite{bsenel2021edgenetsys} that targets the remote clusters using affinities, such as locations and connected devices.\footnote{This is an extension to the \textit{Selective Deployment} mechanism in our previous work~\cite{bsenel2021edgenetsys}. There, workloads could be deployed to remote nodes within a single geographically dispersed cluster. Now, workloads can be deployed to entire remote clusters within a federation of clusters.}
A manager entity, called the \textit{Federation Manager}, is informed by the local cluster for federation-scoped Selective Deployments.
When it receives one, it searches for remote clusters that satisfy the affinities, in order to deploy the workload there on behalf of the tenant.
To move towards a production federation architecture, issues such as caching and scheduling will need to be tackled.

\subsection{Tenant Resource Quota}
\label{sec-architecture-trq}

As described in Sec.~\ref{sec-tenant-resource-quota}, Kubernetes provides the ability to associate resource quotas with namespaces, but in the context of independent namespaces.
Since our multitenancy framework extends Kubernetes namespaces to work in a hierarchical fashion, we need to extend the quota mechanism to take into account the dependency of each namespace on other namespaces above it and below it in the hierarchy.  
The EdgeNet quota mechanism is designed to allow for a given resource to be shared out between a namespace and its child namespaces, and for the parent namespace to recoup each child's portion when it is relinquished. 
Child namespaces can in turn share out their quota with their children, and so on, recursively.
Our framework covers the following resources: CPU, memory, local storage, ephemeral storage, and bandwidth, each accounted for individually.\footnote{Tenant resource quotas will be expanded to include other resources in the future, such as namespaces, pods, and configmaps.}

We model tenant resource quotas by representing the tree of a hierarchical namespace as a graph $T=(V,E)$ composed of vertices $V$ and parent-to-child edges $E$. For our purposes, each vertex $v \in V$ is a namespace, except for the root node. 
The tenant of a namespace $v$ is entitled to construct a subtree $T_v$ rooted at that namespace $v$, which is also called a core namespace.
Denote $q(T)$ the resource quota of tree $T$, and each namespace $v \in V$ has a resource quota $q(v)$.
Here, we assume that there is only a quota for different types of resources for simplicity.
In fact, different quotas can be set for different resources, such as CPU and memory.

Let $\sigma(v) = \{w_1,w_2 \ldots\} \subset V$ represent the subnamespaces of $v$.
Likewise, assume $\sigma(w) = \{z_1,z_2 \ldots\} \subset V$ represent the subnamespaces of $w$.
The hierarchical resource quota problem here is twofold.
First, we must ensure that a tenant resource quota $q(T_v)$ is equal to aggregated resource quota across all its namespaces: $q(v)+\sum_{w \in \sigma(v)}q(w)+\sum_{z \in \sigma(w)}q(z)$.
The latter is to guarantee that the resource quota allocated to a subtree rooted at a namespace $w$ is also equal to aggregated resource quota across the namespaces of that subtree, thus $q(T_{w})=q(w)+\sum_{z \in \sigma(w)}q(z)$.

We solve this problem by partitioning resource quotas among parents and their children while keeping with the container orchestration tool's declarative approach. 
A tenant resource quota works by applying an identical resource quota, a Kubernetes resource, to the tenant’s core namespace.
Then, each subsidiary namespace in the core namespace takes its portion from that resource quota, as shown in Fig.~\ref{fig:resource-quota-edgenet}. %explained in the caption of

As mentioned above, when resources are constrained, ensuring a fair share of them is essential.
Static allocation of quotas, however, may lead to inefficient use of the resources.
There are two sides to this problem.
Such resource quotas that are allocated to tenants, assuming some tenants' resource consumptions are inferior to their quotas, may result in suboptimal utilization of compute resources in clusters.
Likewise, the resource quotas that are allocated statically to subnamespaces by tenants, assuming some subnamespaces consume fewer resources than their quotas, may provoke less-than-ideal use of their tenant resource quotas.
Even though our system allows temporary addition to and removal from tenant resource quotas as well as manually updating subnamespace quotas, this solution cannot scale when there are many clusters.
Sec.~\ref{sec-future-work} introduces how we plan to address this problem.

\subsection{Slice and Slice Claim}
\label{sec-architecture-slice}
Two software entities enable node-level slicing; \textit{slice} and \textit{slice claim}. 
Slice, a cluster-scoped entity, forms a subcluster by slicing among nodes, as its name signifies. 
A slice isolates the nodes within it from multitenant workloads once it is established.
These nodes are chosen via a selector composed of fields that denote labels, number of nodes, and desired resources.
On the other hand, a slice claim is a namespaced entity that tenants may create for their subnamespaces.

Nodes in a slice remain in the pre-reserved status until a subnamespace uses that slice.
Once a subnamespace is bound to a slice, the multitenant workloads that runs on the nodes in this slice are terminated within a grace period of a minute.
That is to say, workloads created in that subnamespace are isolated from other tenants.
Thus, the container runtime configuration within such subnamespaces becomes available to tenants.\footnote{Resource-constrained environments may compel CaaS to operate on bare metal.
We will, therefore, assess the performance of Kata with a specific experiment setup described in Sec.~\ref{sec-future-work}.
}
Regarding the termination grace period, we have set it to one minute by default, as twice the default grace period of 30 seconds in Kubernetes.
However, providers can adjust this termination grace period according to their requirements.

\begin{figure*}[hbtp]
  \centering
  \includegraphics[width=0.7\linewidth]{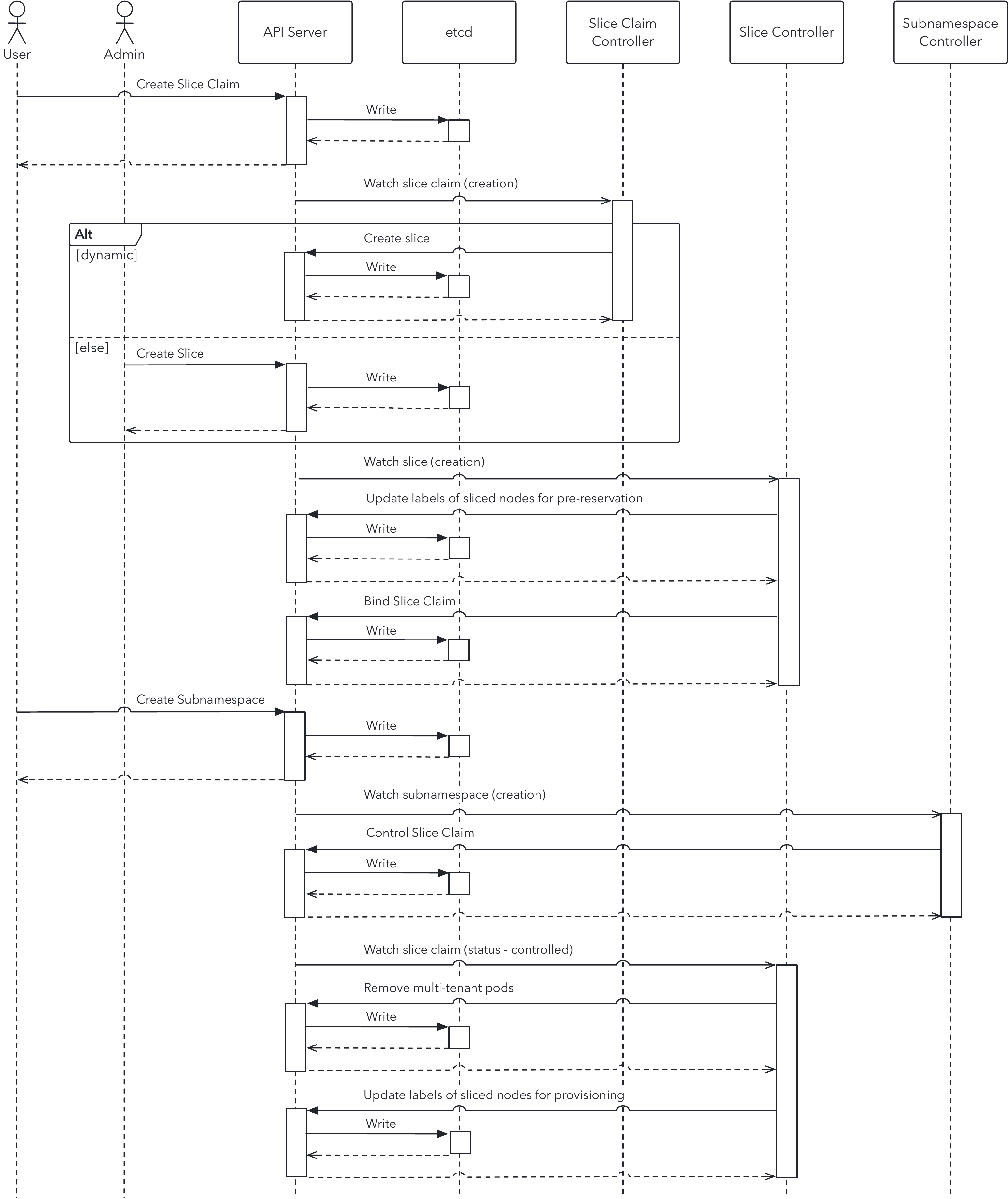}
  \caption{Sequence diagram of a tenant acquiring node-level isolation.
  For the dynamic mode, we assume the tenant has enough quota in the namespace.}
  \label{fig:slice-diagram}
\end{figure*}

A slice claim has two working modes; \textit{dynamic} and \textit{manual}.
The dynamic mode permits a tenant to automatically create a slice if the resource quota in the slice claim's namespace is sufficient.
In contrast, the manual mode prevents a slice claim from generating a slice even if the slice claim's namespace has an adequate resource quota. 
In this case, a cluster administrator must satisfy the tenant's request.
This kind of behavior can be desirable if the number of nodes in a cluster is scarce.
Fig.~\ref{fig:slice-diagram} depicts how a tenant can receive node-level isolation.
We discuss the need for a daemon to improve isolation in Sec.~\ref{sec-future-work}.

\subsection{Admission Control Webhook}
\label{sec-architecture-admission-webhook}
An admission control webhook is a software entity that allows for enforcing custom policies. 
It can mutate and validate object operation requests of users. 
Such mutating and validating operations are critical so as to ensure that users adhere to framework-specific policies.
We enforce custom policies for subnamespaces, slice and slice claim, role requests, tenant requests, cluster role requests, as well as pods.

Kubernetes, by default, lets users pick runtime classes that they desire for their pods.
Likewise, in our framework, a tenant can select the container runtime for the containers running on the nodes in its slice.
However, this is not the preferred behavior unless a tenant acquires entire nodes through node-level slicing.
For the purpose of better isolation, we constantly mutate the runtime class to employ Kata (see Sec.~\ref{sec-container-isolation}) for multitenant workloads through admission control.

\subsection{Role Request}
\label{sec-architecture-rolerequest}
This feature facilitates permission management at namespace scope.
Thanks to its design, this entity provides granular control over tenant users.
A user can request a specific role in a core namespace or any subnamespace of a tenant. 
This role can be one of the cluster roles offered by the cluster provider or a role in that namespace.
Once a request is made, an authorized user in that namespace can approve or deny it.
That is to say, tenant owners and admins can delegate responsibilities to team leaders in child namespaces.
When a tenant represents a large organization, delegation becomes crucial to facilitate management.

\subsection{Other Entities}
\label{sec-architecture-others}
There are two more assistant entities.
A \textit{tenant request} stands for tenant registration. 
A central administrator or a trust strategy, for example, credit card verification, can approve the establishment of tenants and their owners.
In our implementation, a cluster admin may approve a request or deny it.
Another option is, as mentioned above, a provider can integrate a credit card verification-like mechanism with our framework to avoid the manual administration of clusters, supporting CaaS to operate with many clusters at scale.
There are four pieces of information in the request; the organization, the owner, the tenant resource quota, if desired, and whether or not to apply a cluster-level network policy.
A \textit{cluster role request} is an entity that allows a user to claim to hold a role at the cluster scope.
This entity eases shaping a cluster administration team and encourages the platform users to ask for the roles that they need.

\subsection{Authentication}
\label{sec-architecture-authenication}

Our general design approach is to build, wherever possible, upon what is already available for Kubernetes, as we do by adopting OpenID Connect (OIDC)\footnote{OpenID Connect \url{https://openid.net/connect/}} running on top of OAuth 2.0\footnote{OAuth 2.0 \url{https://oauth.net/2/}} as our authentication method.
A feature that is still under development is to extend OIDC with Pinniped\footnote{Pinniped \url{https://pinniped.dev/}} so as to access resources across clusters.
This allows a user to authenticate once to access namespaces and objects, for which the user has access rights, in all of the clusters to which the objects have propagated.

\section{Benchmarking}
\label{sec-benchmarking}
This section analyzes the performance of our EdgeNet single-instance native Kubernetes multitenancy framework.
One of our goals is to assess to what extent native and multi-instance approaches are suitable for edge computing use cases. 
To this end, we compare our framework to single cluster per tenant offerings with the help of Rancher Kubernetes Engine (RKE)\footnote{RKE \url{https://rancher.com/products/rke}} in order to automate cluster creations and to the VirtualCluster~\cite{zheng2021multi} code that realizes a multi-instance-based multitenancy framework. 
That is to say, to represent the multi-instance through multiple clusters approach, we pick RKE, which is widely known for installing Kubernetes; VirtualCluster for the multi-instance through multiple control planes approach, which is a Kubernetes working group framework that is described in the scientific literature~\cite{zheng2021multi}; and our own EdgeNet framework is single-instance.

Both RKE and VirtualCluster perform well when the compute resources are nearly unlimited, or scalability with regard to the number of tenants is less of a concern.
Compared to RKE, VirtualCluster is well-adapted to address the issues of the single cluster per tenant solution, such as high overhead. 
However, as we shall see, there is a tradeoff between performance and isolation, which means that existing solutions are not ideal for edge computing.

We used the \textsc{Geni} infrastructure~\cite{GeniBook2016} to spawn four Ubuntu 20.04 LTS virtual machines with 8 CPUs and 16\,GB of memory in order to conduct experiments with EdgeNet and VirtualCluster.
Using these virtual machines, we created a Kubernetes v1.21.9 cluster consisting of one control plane node and three worker nodes.
The control plane node is completely isolated from any workloads.

For the VirtualCluster experiments, we reserved a worker node for running the manager, syncer, and agent components. 
Likewise, the per-VirtualCluster-tenant entities, which are apiserver, etcd, and controller-manager, are deployed on a dedicated worker node.
For the EdgeNet experiment, an isolated worker node was sufficient to run the entities.
A separate worker node hosted monitoring tools in both cases.
We used the default configuration settings for both frameworks, including the number of workers that process concurrently and the execution period that triggers the controller.

We compared the frameworks' performance for tenant creation and for pod creation.
For VirtualCluster tenant creation, inter-arrival times of 0, 8, 16, and 32 seconds were used for creating 2, 4, 8, 16, 32, and 64 tenants, respectively.
For EdgeNet, inter-arrival times of 0, 2, 4, 8, 16, and 32 seconds were used for creating up to 10,000 tenants.
(We discuss the reasons for the disparity in the number of tenants below.)
For both framework, pods created  were 1,000, 2,500, 5,000, and 10,000.
Timeout is two minutes to create tenants and pods separately.

\begin{figure*}
    \centering
    \subfloat[Symbols represent frameworks, and each inter-arrival time is colored.
    The longer the time between arrivals, the higher the number of successfully created tenants. 
    Stably, VC manages to create around 40 tenants at most when inter-arrival time is set to 32\,s, while EdgeNet reaches 10,000 with 4\,s.]{
        \includegraphics[width=0.3\linewidth]{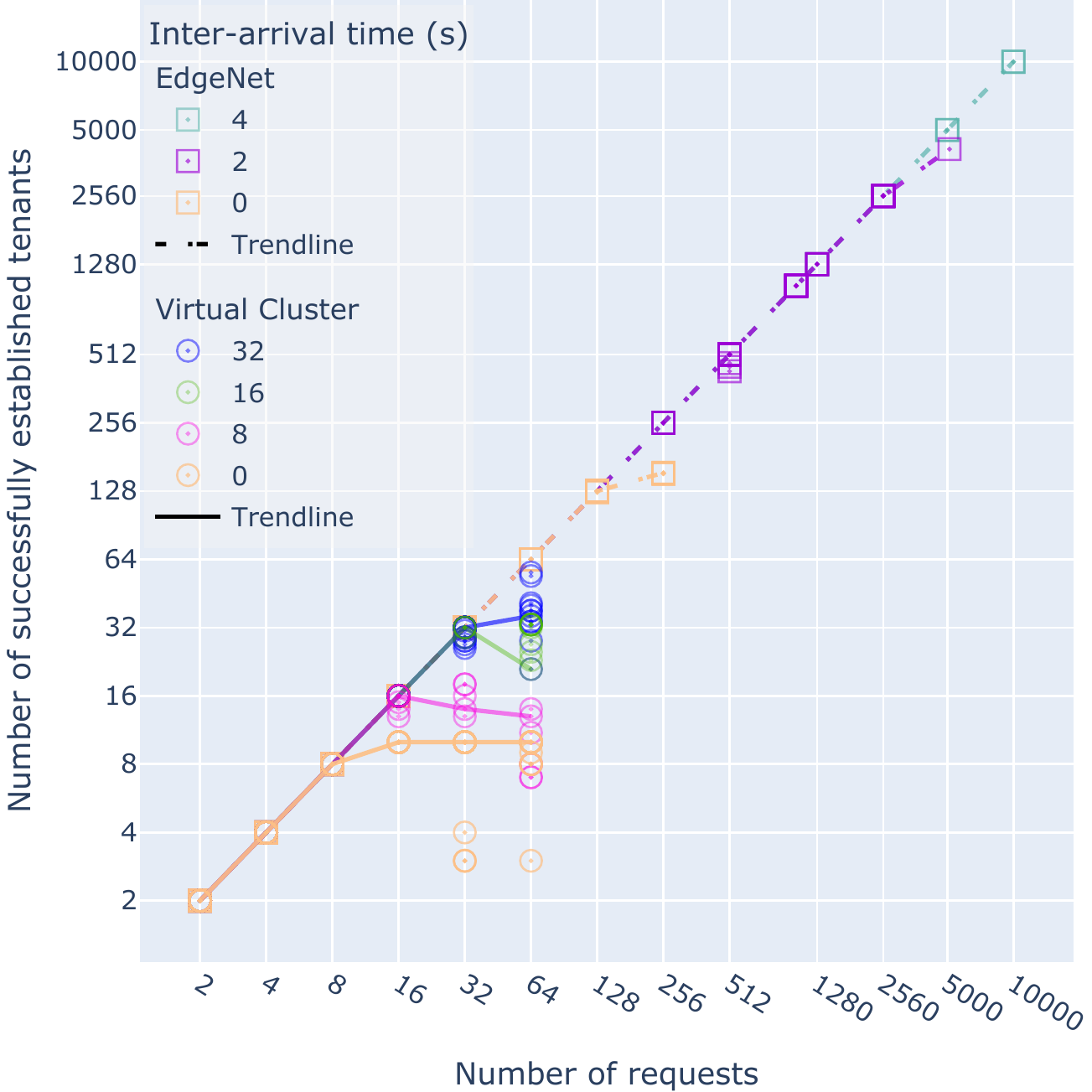}
        \label{fig:vc-edgenet-scatter-successful-creation}
    }
    \hfill
    \subfloat[Box spans from quartile 1 to 3, a line inside indicates the median, and outliers are shown as single points.
    Inter-arrival time affects tenant creation time. 
    The shorter it is, the longer the creation, with higher variation.
    EdgeNet consistently outperforms VC.
    The results are for 32 tenants.
    ]{
        \includegraphics[width=0.3\linewidth]{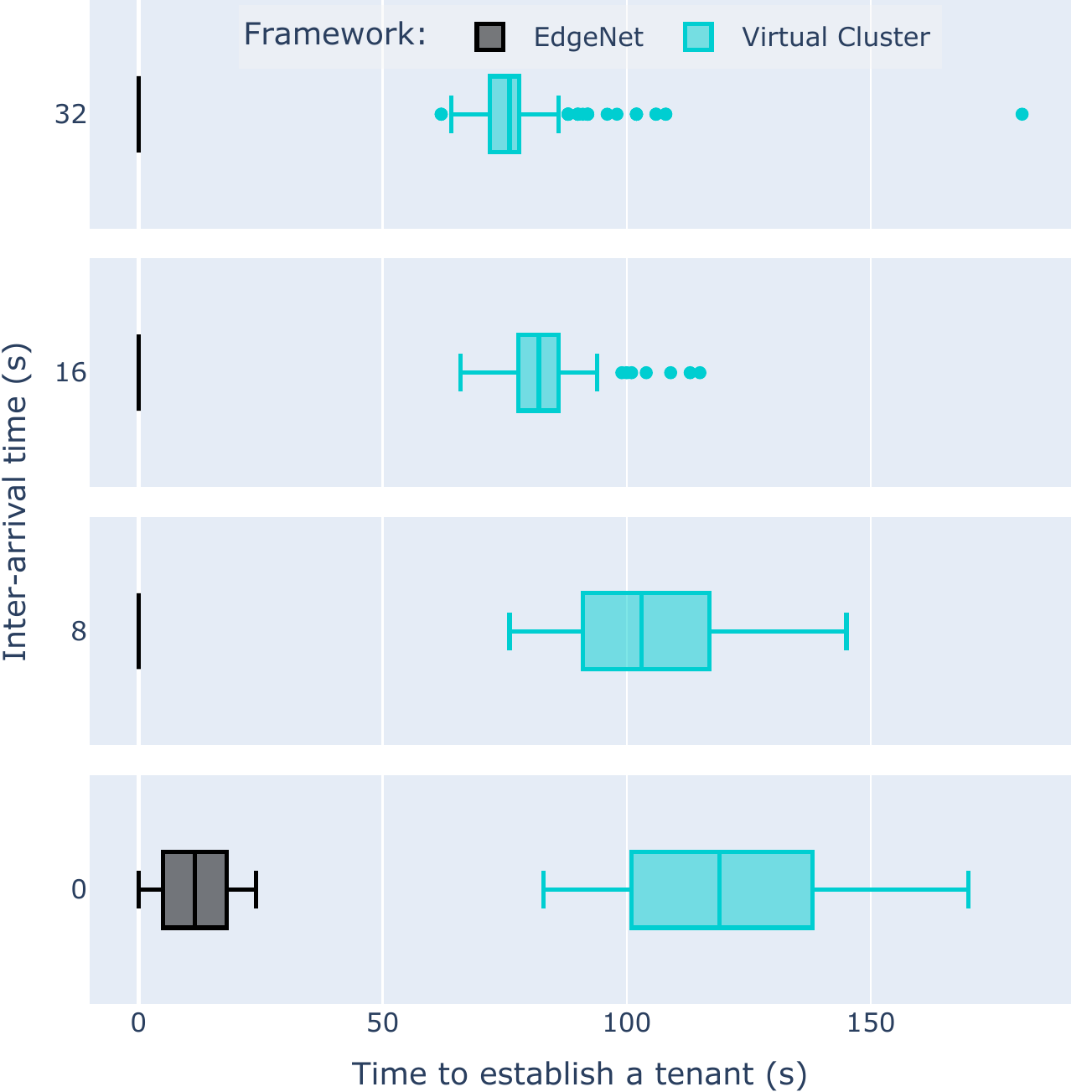}
        \label{fig:vc-edgenet-boxplot-ttct}
    }
    \hfill
    \subfloat[Base resource consumption rises with the number of tenants due to the multi-instance approach in VC. 
    EdgeNet does not introduce such overhead per tenant.
    Resource consumption of framework-specific software entities is insignificant for both.
    No user activity is involved.]{
        \includegraphics[width=0.3\linewidth]{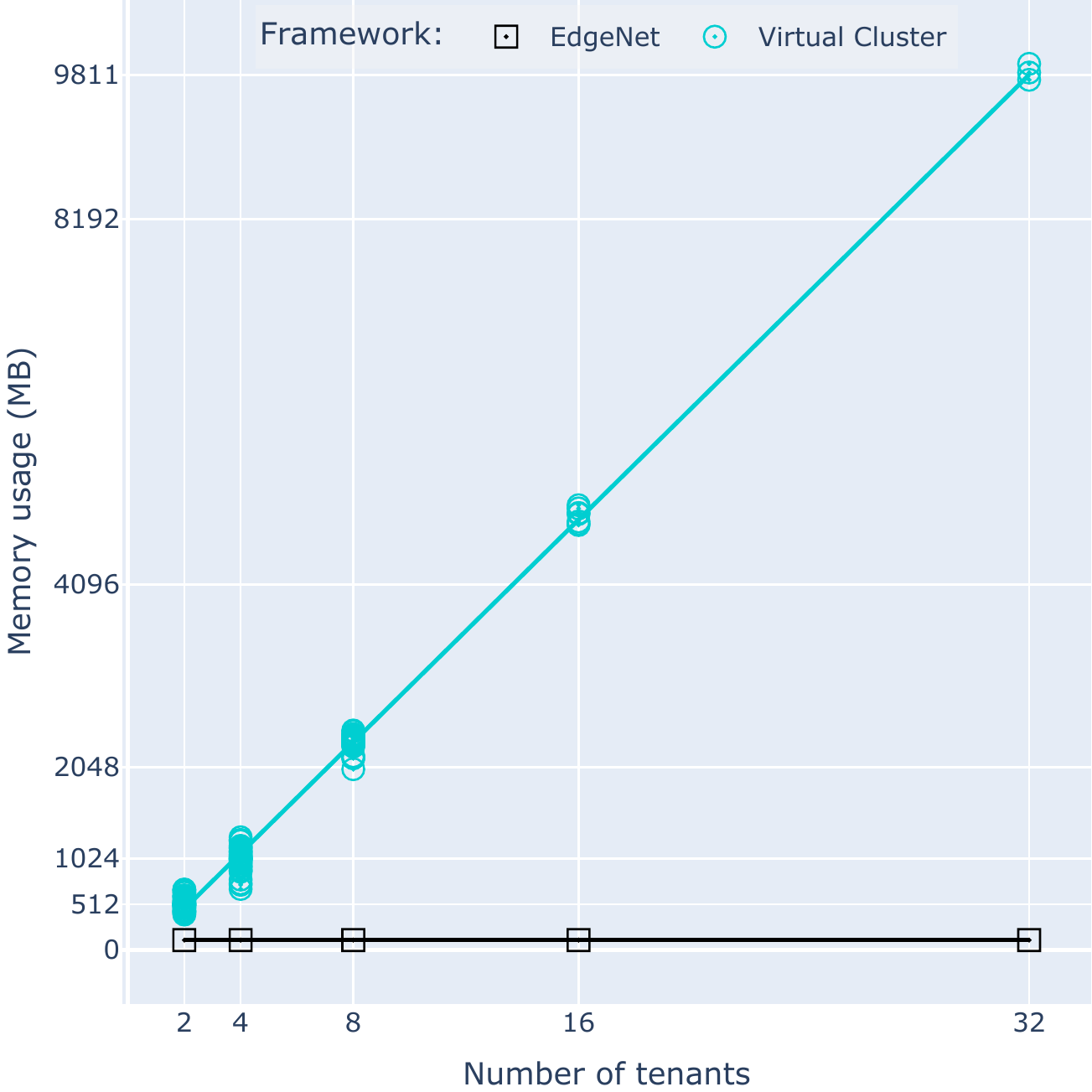}
        \label{fig:vc-edgenet-scatter-memory}
    }
    \caption{Experiment results for VirtualCluster and EdgeNet.}
    \label{fig:experiments}
\end{figure*}

To measure the performance of a cluster per tenant method, reserved resources for tenant entities, a virtual machine with 8 CPUs and 16\,GB of memory, were divided evenly among four Ubuntu 20.04 LTS virtual machines with 2 CPUs and 4\,GB of memory on GENI. 
2 CPUs were chosen because cluster provisioning repeatedly failed with VMs having a single CPU. 
We repeated measurements at least three times for each case.

\subsection{Tenant Creation}
\label{sec-benchmarking-tenant-creation}

As discussed throughout the paper, besides security, overhead is a noteworthy factor in qualifying a multitenancy framework, especially for edge clouds.
Our experiments measure a framework implementation's ability to handle simultaneous creation requests; the time it takes to create a tenant; entities' resource consumption; and consumption per tenant, if it exists.
Each request is considered successful if the framework returns a success status within two minutes after the control plane receives the request.

\subsubsection{VirtualCluster} 

The experiments show a correlation between request inter-arrival time and tenant creation success rate.
For example, with a 32\,s inter-arrival time for 32 creation requests, the number of successfully created tenants ranges from 26 to 32; 
when the inter-arrival time is lowered to 8\,s, the successes decrease to between 13 and 18, as shown in Fig.~\ref{fig:vc-edgenet-scatter-successful-creation}.
It is possible that VirtualCluster's difficulties in handling simultaneous requests stem from an implementation issue that starves tenants of the compute resources necessary to establish their control planes in these circumstances.

Similarly, as seen in Fig.~\ref{fig:vc-edgenet-boxplot-ttct}, decreasing the request inter-arrival time increases the tenant creation time.
At a 32\,s inter-arrival time, the median creation time is 76\,s;
put another way, it would take more than an hour to create 128 tenants.
Furthermore, as the figure shows, the creation time fluctuates more widely as inter-arrival time decreases.

The most critical scaling weakness for VirtualCluster is that every tenant introduces additional overhead in terms of memory and CPU usage due to the per tenant isolation of control plane components: apiserver, etcd, and controller manager.
Fig.~\ref{fig:vc-edgenet-scatter-memory} presents the regular memory usage for 2, 4, 8, 16, and 32 tenants.
For example, a thousand tenants would consume around 300\,GB of memory just to be present in the cluster.
This limitation ultimately affected our experiment, which could not reach a high number of tenants on the single node that we had reserved for tenant components; 
the maximum number of tenants that we could create stably was approximately 40.

In addition to this, a tenant starting to use the cluster results in an increase in resource consumption.
We also noticed that a successful status message for the tenant control plane does not imply that all its components are present and functioning properly.
Therefore, we only considered the cases where control plane components per tenant were all created successfully.

\subsubsection{EdgeNet} 
\label{sec-benchmarking-tenant-creation-EdgeNet}

As opposed to VirtualCluster, EdgeNet supported the creation of 128 tenants simultaneously with an almost zero failure rate across experiments.
It also scaled well beyond this number, stably generating 2,560 and 10,000 tenants when the request inter-arrival time was set to 2\,s and 4\,s respectively, as shown in Fig.~\ref{fig:vc-edgenet-scatter-successful-creation}.
This is as far as one can go before running into Kubernetes' maximum namespace threshold\footnote{Kubernetes Scalability SIG documentation: \textit{Kubernetes Scalability thresholds} \url{https://github.com/kubernetes/community/blob/master/sig-scalability/configs-and-limits/thresholds.md}} of 10,000 in a cluster;
if tenants are allowed to have around ten namespaces each, the number of tenants per cluster is limited to around 1,000.

When requests arrive simultaneously, the median time for EdgeNet to create a tenant object in the control plane increases with the number of tenants: 38\,ms, 48\,ms, 63\,ms, 68\,ms, 106\,ms, 175\,ms, 216\,ms, and 270\,ms for 2, 4, 8, 16, 32, 64, 128, 256 tenants respectively.
Another pattern of results is obtained with an inter-arrival time of 2\,s: creation times are 11\,ms for 1,280 tenants, 11\,ms for 2,560 tenants, and we tested as far as 5,120 tenants, also clocking in at a median of 12\,ms.
For 10,000 tenants, the median value is still 12\,ms when inter-arrival time is set to 4\,s.
However, the maximum values increase as a function of the number of requests.

This suggests that concurrent or many requests saturate the shared API server, controller manager, and etcd moderately.
Thus, when arrivals are simultaneous, the average time to fully establish a tenant increases as follows: 500\,ms for 2 tenants, going up to 937\,ms for 128 tenants.
But Fig.~\ref{fig:vc-edgenet-boxplot-ttct} reveals that the time to fully establish a tenant drops when requests are spread out in time.
For 32 tenants, the median times are 11.5\,s for simultaneous arrivals, 271\,ms for 8\,s, 274\,ms for 16\,s, and 274\,ms for 32\,s.

Good results are seen for EdgeNet since it configures the state of the cluster rather than replicating the components, it does not generate per-tenant overhead, as shown in Fig.~\ref{fig:vc-edgenet-scatter-memory}.
Given that the resource consumption of controllers is negligible, it is fair to state that there is no significant overhead in our framework.

It takes EdgeNet approximately 1\,min\,41\,s to create 128 tenants.
Furthermore, EdgeNet's creation time can be shortened if needed by adjusting the number of workers and the running period.
By default, the tenant controller uses two workers with a running period of 1\,s, and the client's query per second (QPS) rate and burst size are set to 5 and 10, respectively.
We tried altering the setup to have ten workers with a 500\,ms  running period, setting QPS and burst to 1,000,000 each.
With these settings, it takes just 17\,s to fully create 128 tenants, as seen in Fig.~\ref{fig:edgenet-comparison-simultaneous}.
The same figure shows that EdgeNet can handle simultaneous requests if a cluster welcomes around 1,000 tenants.
The time it takes to establish all tenants eventually converges towards two minutes for both settings, thereby satisfying the success criteria we described at the beginning of Sec.~\ref{sec-benchmarking-tenant-creation}. 
However, we noticed it surpasses two minutes when simultaneous requests are more than 1,280.
We presume that this may be due to client or control plane saturation resulting in the API server receiving delayed requests, which we need to investigate further.
Fig.~\ref{fig:edgenet-comparison-inter-arrival} shows that EdgeNet with default settings can scale up to 10,000 tenants when inter-arrival time is set to 4\,s, but it takes more than ten hours in total.

\subsubsection{Comparison} 

Our findings on tenant creation at least hint that better isolation provided by the multi-instance approach comes at the cost of performance loss.
What can be clearly seen is that EdgeNet surpasses VirtualCluster on scalability and speed.
The peak number of tenants in a cluster is 10,000 for EdgeNet but around 40 for VirtualCluster, even with longer inter-arrival times.
VirtualCluster offers a separate control plane per tenant, meaning an increase in base resource consumption, which is one of the major limitations.
In contrast, EdgeNet can scale up to the cluster namespace threshold thanks to the native approach discussed in Sec.~\ref{sec-multitenancy-approach}.

\begin{figure}
    \centering
    % Simultaneous requests for both cases.
    \subfloat[Default settings vs. optimization.]{
        \includegraphics[width=0.45\linewidth]{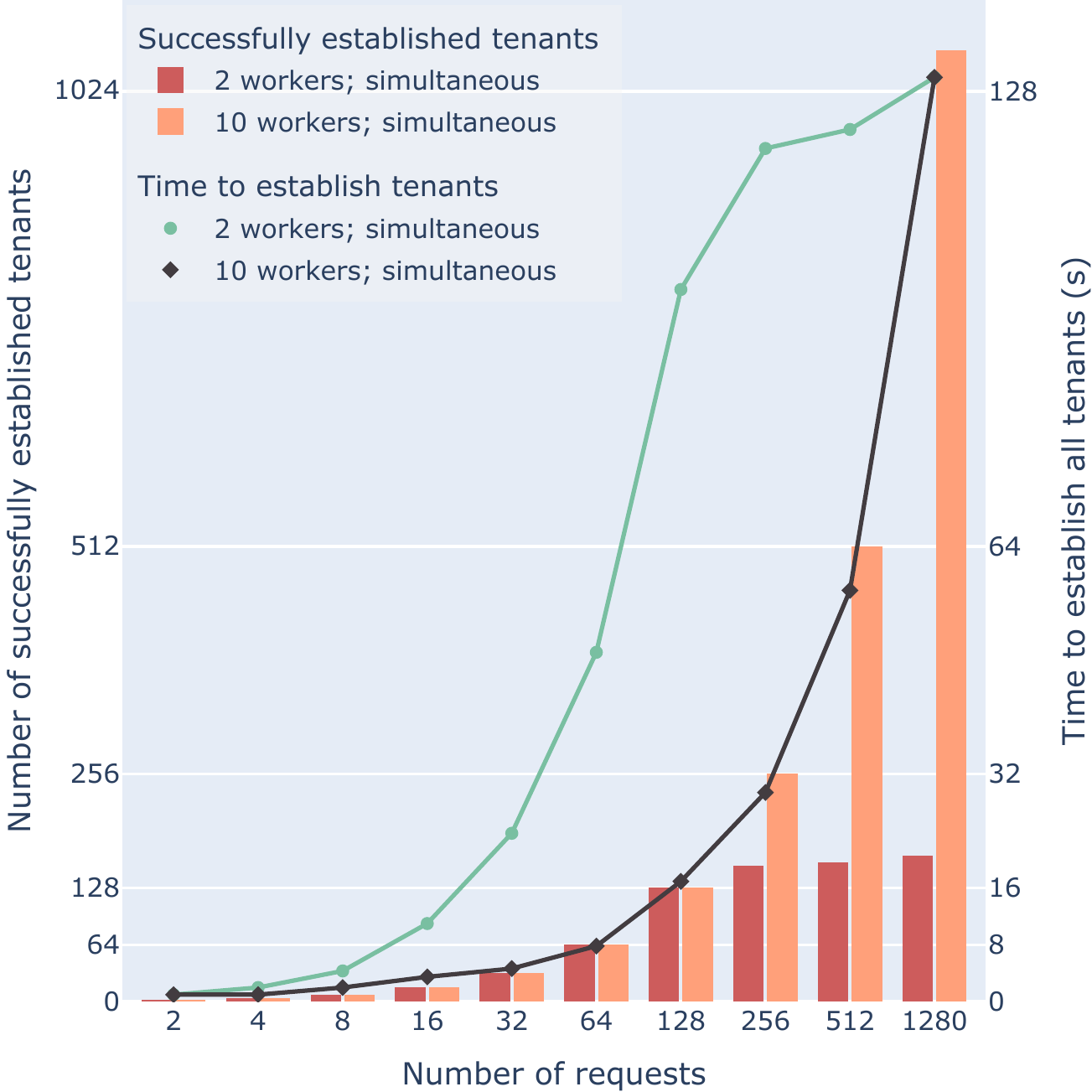}
        \label{fig:edgenet-comparison-simultaneous}
    }
    \hfill
    % 2 workers: arrivals are separated by 4\,s, 10 workers: simultaneous.
    \subfloat[Inter-arrival time vs. optimization.]{
        \includegraphics[width=0.45\linewidth]{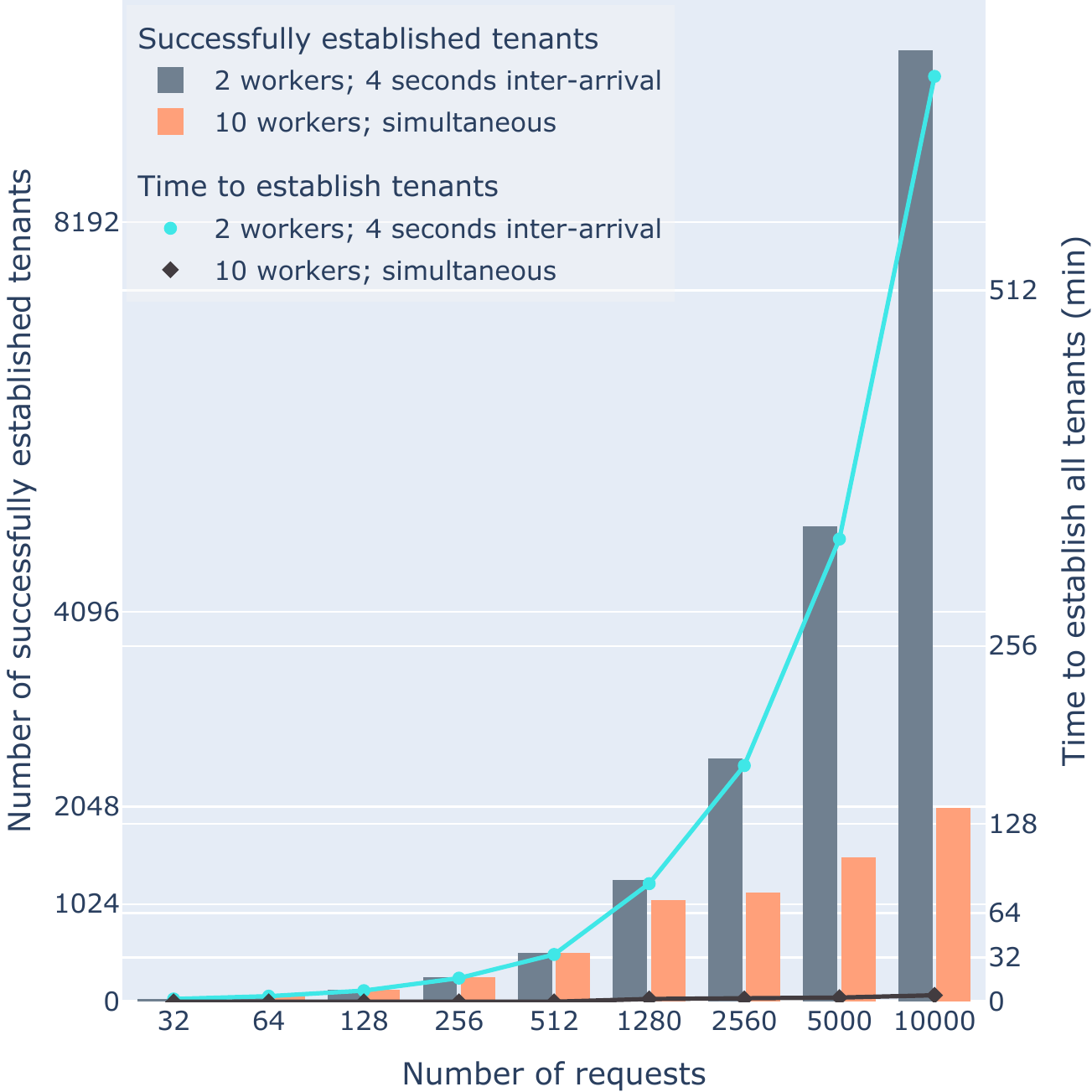}
        \label{fig:edgenet-comparison-inter-arrival}
    }
    \caption{Effect of the number of workers, running period, QPS and burst on EdgeNet's performance, including comparison with the effect of time between arrivals.}
    \label{fig:edgenet-comparison}
\end{figure}

Scalability is only one aspect of evaluating a framework's performance, especially for edge-specific workloads. 
Speed, stability, and overall reliability are also important.
EdgeNet is considerably faster than VirtualCluster at tenant establishment for all inter-arrival times.
Fig.~\ref{fig:edgenet-comparison-simultaneous} shows how optimizing the number of workers, running period, QPS, and burst can further improve EdgeNet's performance.
Furthermore, when arrivals are not simultaneous, EdgeNet handles each request in microseconds, whereas VirtualCluster takes seconds, even minutes.
Speed is an important contributing factor to establishing many tenants concurrently or in sequence, but stability and reliability are also critical.

VirtualCluster cannot adequately address simultaneous requests or requests with a short inter-arrival time, even if they are not many. 
Because of this issue, we observe a marked fall in the success rate of tenant establishment in such cases.
We speculate that an implementation issue might be provoking resource starvation in tenant control planes.
The time it takes to finish establishing all tenants is significantly more deterministic for EdgeNet than for VirtualCluster; EdgeNet exhibits almost no variation, irrespective of whether 128 tenants or 2,560 tenants are being created.
However, EdgeNet’s performance is tied to the control plane capacity as well.  
When many requests with little time between arrivals oversaturate the control plane, it has difficulty establishing all tenants properly.
Nonetheless, EdgeNet can process 1,000 simultaneous requests, allowing tenants to use ten namespaces for each, as discussed above.

The multi-instance approach limits VirtualCluster's scalability since the base resource consumption increases as tenant numbers grow; providing one control plane per tenant costs about 285 MB of memory each.
It is a large memory consumption, especially for edge computing use cases.
This would also increase cloud computing expenses per tenant.
Such overhead is not present in EdgeNet.

VirtualCluster, with its multi-instance approach, falls on the isolation side of the isolation-performance trade-off.
Thus, performance degradation is expected.
Considering this information is crucial to interpreting results correctly.
Table~\ref{tab:quick-comparison} shows how much better VirtualCluster performs than a single cluster per tenant system.
Regardless, our framework produced a more robust outcome with a significant performance advantage.
Based on the results, EdgeNet is better suited for edge computing, as well as for cloud-edge collaborations.

\subsection{Pod Creation}

To examine the effect of VirtualCluster's syncer on performance, we measure the time that it takes to create a representation of a pod as an object.
The syncer gathers pod objects from the tenant control plane and creates them in the host control plane, called a \textit{supercluster}.
The time that we measure is the time that it takes for  pods to show a pending status in the supercluster.

We focus on pods instead of containers as that helps us reveal the framework's capabilities, as the creation of containers is dependent upon many factors such as available resources on the host, container runtime, and volume type.
There are already many papers that evaluate these aspects for different container runtimes.

\begin{table}[t]
\centering
\caption{Quick comparison of native and multi-instance approaches}
\label{tab:quick-comparison}
\begin{threeparttable}
\begin{tabular*}{\linewidth}{l@{\extracolsep{\fill}}ccccc}
    \toprule
    \multirow{2}{*}{} &
    \multirow{2}{*}{Max Tenants} &
    \multicolumn{2}{c}{Creation Time (s)} & 
    Per Tenant
    \\
    \cmidrule(lr){3-4}
    & & Median & Max & Overhead (MB) \\
    \midrule
    EdgeNet & 10,000 & 0.616 & 1.741 & None \\[0.5pt]
    VC & 40 & 82 & 93 & 285 \\[0.5pt]
    RKE & 4 & 343 & 395 & 711 \\[0.5pt]
    \bottomrule
\end{tabular*}
\begin{tablenotes}
  \item Max number represents the maximum number of successfully established tenants that can be stably reached with respect to allocated resources for tenant creation.
  \item The time it takes to establish a tenant for four simultaneous requests.
  \item Per tenant overhead refers to the fixed proportion of resources each tenant consumes in an average manner, regardless of activity. 
  VC consumption was measured using pods that deliver control planes to tenants, and RKE consumption was measured through containers that provide clusters to tenants.
  Traditional VM-based overhead is not included in RKE.
\end{tablenotes}
\end{threeparttable}
\end{table}

\subsubsection{VirtualCluster} 

Request inter-arrival time affects the number of successfully created pods similarly to how it affects the number of successfully created tenants.
This is because, as described above, created tenants struggle to enter a healthy state when inter-arrival time is short.
Also, increasing the number of tenants that create pods degrades pod creation performance.
One reason for this is the increased resource use in tenants' control planes.
Pod creation success, for 16 or 32 tenants, is 100\% for 1,250 and 2,500 pods;
performance drops slightly for the creation of 5,000 pods; for 10,000 pods, a median of 9,431 are successfully created.
The time it takes to create pods also increases with the number of pods created.

\subsubsection{EdgeNet} 

Up to 10,000 pods can be created simultaneously for up to 300 tenants in a deterministic manner.
Performance for 10,000 pods started to degrade at 384 tenants due to saturation of the Kubernetes API server in processing requests from different core namespaces. 
Pod creation time does not spike, as there is no intermediate layer syncing the objects.
A linear relationship is observed between the median creation times and the number of pods in Table~\ref{tab:time-to-create-pod}.

\subsubsection{Comparison} 

In VirtualCluster, the syncer is an intermediate layer between the supercluster and tenant control planes in order to sync pod objects.
The disadvantage of this approach is that every pod operation introduces synchronization overhead, both on the supercluster and tenant sides.
We should emphasize that every synchronization process causes a delay for a pod to be up and running.
This may raise concerns about running VirtualCluster at scale; however, it can be mostly overcome by providing more computing resources to the framework, leading to higher costs.
In contrast, EdgeNet allows tenants to directly make use of the same control plane so as to create pods.
Its performance is directly related to the capabilities of the control plane.
Thus, EdgeNet produces superior results, where VirtualCluster takes at least three times as much time as EdgeNet to create 1,250 pods, 2,500 pods, 5,000 pods, and 10,000 separately.
Table~\ref{tab:time-to-create-pod} shows how far VirtualCluster's synchronization of objects between the supercluster and tenant control planes causes significant delays while achieving better isolation.

\begin{table}[htbp]
\centering
\caption{Time in seconds, median values, to create a representation of a pod as an object in the host control plane. The number of tenants used for the experiments is set to 32 for both VirtualCluster and EdgeNet}
\label{tab:time-to-create-pod}
\begin{threeparttable}
\begin{tabular*}{\linewidth}{l@{\extracolsep{\fill}}c|c|c|c}
    \toprule
    Number of pods & 1,250 & 2,500 & 5,000 & 10,000 \\
    \midrule
    VirtualCluster & 12 & 22.5 & 46 & 134 \\
    EdgeNet & 2.5 & 6 & 12 & 27 \\
    \bottomrule
\end{tabular*}
\end{threeparttable}
\end{table}

\section{Future Work}
\label{sec-future-work}
Although the work presented in this paper goes a long way to establishing a Kubernetes multitenancy framework that is suitable for the edge cloud, there is still considerable room for improvement.
We describe areas for future work below.

\textbf{Resource Quota Optimization.} We plan to develop an optimization algorithm that distributes, in a best-effort fashion, underutilized tenant resource quotas among the ones who consume all of their quotas and surplus subnamespace resource quotas among those in the same tenant who hit their quotas.

\textbf{Sub-node-level VIP Slicing.} In order for tenants to receive guaranteed access to resources that are both available and dedicated to them, node-level slicing is currently the only option.
By adding a new point to the slice spectrum, it will be possible to do so at sub-node-level granularity.
We will deploy a pod that consumes almost no real resources on a node to ensure that resources are secured.
Priority classes will enable the reservation mechanism for pods.

\textbf{Storage.} Sharing storage among containers securely at the edge is a challenge due to the security issues discussed in the Rationale section (see Sec.~\ref{sec-rationale-security-and-performance}).
We plan to develop an agent that runs on every node and is ready, upon tenant demand, to prepare a disk partition that the tenant can use as a storage volume for its Kata containers. 

\textbf{Security.} We plan to encrypt each tenant's data separately, across its namespaces and cluster-scoped resources.
In this way, even if a tenant's data leaks, another tenant will not be able to read it.

\textbf{Customization.} Tenants cannot currently create cluster-scoped resources independently.
We plan to develop a namespace-scoped custom resource that allows users to dynamically create cluster-scoped resources.
This entity will be using the namespace name as a prefix in generating cluster-scoped resource names to avoid collisions.

\textbf{Subnamespaces.} A user may want to attach labels to subnamespaces. 
There is a risk, however, of a malicious actor breaching another tenant's network policies if labels are defined independently. 
For example, one can launch a brute-force attack to correctly guess the namespace labels used in a tenant's network policies.
By using the name of the subsidiary namespace as a prefix, we plan to solve this issue. 
Inheritance will then allow labels to be passed down from parent to child.

\textbf{Container Isolation.} Based on the reasons outlined at the end of our discussion of lightweight hardware virtualization (see Sec.~\ref{sec-container-isolation}), we will use a specific experiment setup to assess how Kata, gVisor,\footnote{gVisor \url{https://gvisor.dev/}} and runC perform.
We will examine a setup in which Kata and gVisor run on a physical server while runC runs on a virtual machine created on that server.

\textbf{Federation.} Containerized workloads need to move between edge clouds and clouds seamlessly without any user intervention.
By leveraging local authorities such as hierarchical federation managers, we aim to address issues of clusters trusting one another.
We will develop a throughgoing federation architecture and will implement a fully-developed federation framework that works in concert with our multitenancy framework.
Once the implementation is done, we will assess its performance.

\textbf{Isolation Daemon.} Kubernetes garbage collection removes unused images.
However, our slicing feature provides on-demand node-level isolation, so we need to instantly clean the node from multi-tenant pods and container images. 
We also consider clearing up \textit{iptables} rules during this process.
An isolation daemon that runs on each node will be further developed to fulfill these operations.

\textbf{Additional Experiments.} Due to time and resource constraints, we compare a limited number of systems in the benchmarking section.
Likewise, some variables of interest could not be studied.
For the generalizability of our findings, we will conduct additional measurements addressing these two limitations.

\section{Conclusion}
We have presented EdgeNet, a Kubernetes-based multitenancy framework for Containers as a Service (CaaS) that, because it is native, i.e., serves all tenants through a single control plane and a single data plane per cluster, is a more efficient alternative to the current multi-instance manner in which cloud providers offer CaaS.
Our benchmarking results demonstrated good scalability and response times for EdgeNet as compared to a leading multi-instance alternative.
Though, in our framework, tenants are not isolated into separate control planes, their containers nonetheless receive the high level of isolation that is provided by Kata containers.
For edge computing to succeed, we believe that security and isolation must be handled natively in software so that workloads can be moved between distant clusters within short delays.

There are, of course, still many questions to be answered.
What are the most optimal ways to establish a robust CaaS federation that is composed of ubiquitous clusters offered by numerous providers?
In order for clusters to join and leave such a federation seamlessly and securely, what trust mechanisms must be in place?
How can users get reliable and transparent billing systems in such an environment?% , perhaps, with a blockchain-based solution?

Anyone may avail themselves of our liberally-licensed, free, open-source code to enable multitenancy in a Kubernetes cluster.
It is already in production use in the EdgeNet edge cloud testbed, for which the tenants are research groups around the world.
And it is particularly suited for edge clouds, where resources are limited, as well as for the cloud.
Because of its federation features, we see this framework as paving the way for tenants to deploy their services across edge clouds operated by many different operators worldwide.

\section{Acknowledgements}
EdgeNet got its start thanks to an NSF EAGER grant, and now benefits from VMware Academic Program grants via CAF America and the Fondation Sorbonne Université, as well as a French Ministry of Armed Forces cybersecurity grant.

\bibliographystyle{plain}

\end{document}